\documentclass[12pt]{article}%
\usepackage{amsmath}
\usepackage{amsfonts}
\usepackage{amssymb}
\usepackage{graphicx}
\usepackage{xcolor}
\usepackage{color}
\usepackage{geometry}
\usepackage{float}
\usepackage{natbib}%
\usepackage{xurl}
\providecommand{\U}[1]{\protect\rule{.1in}{.1in}}

\numberwithin{equation}{section}

\numberwithin{equation}{section}

\newcommand{\be}{\begin{equation}}
\newcommand{\ee}{\end{equation}}

\newcommand{\bq}{\begin{eqnarray}}
\newcommand{\eq}{\end{eqnarray}}

\setcitestyle{authoryear,round}
\begin{document}
\title{Strategic Investment to Mitigate Transition Risks\footnote{*We thank  
Marcel Oestreich, Amr ElAlfy, Chengguo Weng, and Johnny Li for comments and suggestions. The usual disclaimer applies.
}}
\author{Jiayue Zhang\thanks{Department of Statistics \& Actuarial Science, University of Waterloo (j857zhan@uwaterloo.ca)} 
\and Tony S. Wirjanto \thanks{Department of Statistics \& Actuarial Science, University of Waterloo (twirjanto@uwaterloo.ca)} \thanks{School of Accounting \& Finance, University of Waterloo (twirjanto@uwaterloo.ca)}
\and 
Lysa Porth \thanks{Gordon S. Lang School of Business and Economics, University of Guelph (lporth@uoguelph.ca)}
\and Ken Seng Tan\thanks{Division of Banking and Finance, Nanyang Technological University (kenseng.tan@ntu.edu.sg)}
}
\date{{\small July 2025 }}
\maketitle
\begin{abstract}
\noindent This paper investigates strategic investments to mitigate transition risks in sectors mostly impacted by the shift to a low-carbon economy. It emphasizes the importance of tailored sector-specific strategies and the role of government interventions in shaping corporate behavior. Using a stylized, multi-period model, this paper evaluates the economic and operational trade-offs companies face under four decarbonization scenarios. The analysis provides practical insights for both policymakers and business leaders, demonstrating how regulatory frameworks and strategic investments can be aligned to manage transition risks while optimizing long-term sustainability. 



\end{abstract}

\makeatletter

\makeatother
\baselineskip 18pt

\newpage

\section{Introduction}


Mitigating climate change requires an earnest effort of decarbonization of the global economy. 
Achieving decarbonization demands transformative changes across sectors of industries in the economy. While sectors such as a power generation have made strides toward a renewable energy adoption, industries such as a steel and aviation lag behind in this effort due to underdeveloped low-carbon alternatives (\cite{kober2020global}, \cite{dietz2020tpi}). These challenges highlight the importance of understanding climate risks, which fall into two categories: physical and transition risks. Physical risks encompass direct impacts of climate change, such as extreme weather events, while transition risks arise from shifting to a low-carbon economy driven by regulations, technologies, and societal demands (\cite{carney2015breaking}). This paper focuses its analysis on transition risks as they offer a feasible scope for a strategic intervention and remain relatively underexplored in the
field of quantitative modeling in sustainable finance (\cite{cisagara2024finance}).


Transition risks result from structural changes in industries due to a shift toward a low-carbon economy. High-emission sectors, such as steel, aviation, and energy-intensive manufacturing, face challenges in adapting to stricter regulations and cleaner technologies. Policy uncertainty adds complexity, creating financial risks for firms and investors (\cite{holden2018climate}). These risks include stranded assets, increased credit risk, and volatility in asset returns (\cite{semieniuk2021low}).

While low-carbon policies, such as carbon taxes and emissions trading schemes, are essential for meeting climate objectives, their implementation introduces non-trivial transitional challenges to companies (\cite{boe2018transition}). Companies are evaluated not only on their financial metrics but also on their resilience to climate risks (\cite{schulte2018company}). Addressing these risks effectively requires targeted strategies and robust quantitative models.


The shift toward a low-carbon economy offers opportunities to firms, such as cost savings, market leadership, and innovation, alongside major challenges. Companies like Tesla and BP have successfully aligned with low-carbon strategies, reaping benefits such as new revenue streams and enhanced competitiveness. However, this effort of transition is capital-intensive, with high upfront costs and regulatory uncertainty posing risks to companies (\cite{ngfs2019call}, \cite{schulte2018company}). Additionally, the rapid pace of technological advancements can render current investments obsolete, further complicating any long-term planning that companies are undertaking.

Policymakers and businesses must balance a climate action with financial stability. In the next section, we introduce a highly stylized model to guide low-carbon transition strategies, helping industries navigate uncertainties and optimize their transformational efforts.

\section{Preliminary Steps toward a Quantitative Model}

\subsection{Motivations}


Different industries face different and unique challenges and investment needs in the low-carbon transition. Key sectors in the economy include:
(i) Energy, where transitioning to renewable energy requires upgrading infrastructure and advancing storage technologies (\cite{boyd2013public});
(ii) Transportation, where decarbonization efforts focus on electric vehicles, hydrogen fuel, and charging networks (\cite{tyfield2013transportation});
(iii) Manufacturing, where industries such as steel and chemical processing are exploring low-carbon processes such as hydrogen-based production and carbon capture technologies; and (iv) Agriculture, where investments in regenerative practices and ecosystem preservation are critical for reducing emissions (\cite{van2021cash}).

Transition costs are substantial to firms. Estimates suggest that a global spending on net-zero initiatives could reach \$275 trillion by 2050, with varying investment levels across sectors (\cite{mckinsey2021netzero}). Therefore, tailored, sector-specific strategies are called for to balance costs and achieve decarbonization goals more effectively.


Investing in the low-carbon transition confers great benefits across various industries by driving long-term sustainability goal, reducing regulatory risks, and enhancing competitive positioning. Reducing carbon emissions can lead to cost savings through improved energy efficiency and reduced fuel expenses for energy-intensive industries such as manufacturing and transportation. In the financial sector, supporting low-carbon initiatives can mitigate portfolio risks linked to climate change, while at the same time also opening new investment opportunities in green technologies and sustainable businesses. Industries such as real estate and construction benefit by developing energy-efficient buildings that attract environmentally conscious tenants, which boost property values. Additionally, adopting low-carbon strategies can enhance the brand reputation for corporate responsibility and environmental stewardship, meet evolving consumer preferences for sustainability, and ensure compliance with future climate regulations, positioning companies for growth in a low-carbon economy. However, the upfront costs of adopting cleaner technologies can strain businesses, particularly if the returns on investment are not materialized immediately. For example, investing in renewable energy infrastructure may tie up capital that could otherwise be used for business expansion or other needs.

Companies must carefully balance the costs and benefits of low-carbon investments. This involves determining the optimal investment ratio, which refers to the proportion of financial resources allocated to low-carbon initiatives relative to total investments. Companies that can effectively manage financial risks and seize the long-term benefits of low-carbon investments will likely be able to strengthen their competitiveness and resilience in a rapidly evolving global economy.

Given these considerations, this paper proposes a highly stylized model to determine the optimal low-carbon transition strategy across different sectors of the indutries in the economy. The model aims at capturing the unique underlying dynamics and investment needs of each sector, helping industries navigate the complexities of transitioning toward a sustainable, low-carbon future.

\subsection{Components of the Model}

Variables are introduced in this model to replicate the dynamics of low-carbon transition investments and their effects on various industry operations, offering insights into investment strategies. Below, we provide a detailed description of each of the variables and parameters of the model.

\textbf{Selling price per unit ($p$)}. This variable represents the price at which the product of interest is sold in the market. Taking the selling price into consideration is to describe and quantify the revenue generated by the firm. In the context of low-carbon transition, businesses may find opportunities to increase prices for eco-friendly products due to an increased consumer demand for sustainability. For instance, Tesla's Model 3 and other EVs are often priced higher than traditional internal combustion engine vehicles (ICEs), yet some consumers are still willing to pay the premiums for the environmental advantages.

\textbf{Production Cost per unit ($c$)}. This variable includes all expenses incurred by a company in manufacturing each unit, such as raw materials, labor, and energy. Analyzing production costs is essential to evaluating profitability of the company. A study by \cite{siddique2018unilever} found that switching to sustainable sourcing practices initially raised production costs for the company due to the adoption of cleaner and more expensive technologies, but long-term benefits such as energy savings and brand value will eventually outweigh the initial investment. 

\textbf{Selling Profit per unit ($p - c$)}. This variable reflects the profit earned from each unit sold by a company. It is determined by subtracting the production costs from the selling price. It captures the direct profitability of the product after accounting for costs. Profit is the primary goal for a company when making decisions, as it ensures financial viability and supports long-term growth, influencing how much of the company's revenue can be allocated toward investments, including those for sustainable transitions.

\textbf{Total Asset ($A$)}. This variable represents all financial resources available to a company, encompassing both current and long-term assets. This metric is essential for evaluating the company's capacity to invest in production and growth opportunities, providing insight into the scale and potential of the business. Research indicates that firms with larger asset bases are more inclined to invest more substantially in sustainability initiatives, as they can support longer investment horizons and absorb higher upfront costs associated with such projects (\cite{eccles2014impact}). For example, Amazon has invested \$2 billion in renewable energy through its Climate Pledge Fund, leveraging its extensive financial resources to lead in corporate sustainability (\cite{amazon2020sustainability}). Consequently, total assets not only reflect a company's financial strength but also its commitment to sustainable practices and long-term value creation. 

\textbf{Transition Investment Ratio ($\alpha$)}. This variable is a critical metric in our model. It measures the proportion of total assets allocated to low-carbon initiatives. This ratio is essential to a firm as it reflects its commitment to sustainability and its responsiveness to market dynamics and regulatory pressures. By strategically adjusting this ratio, businesses can enhance their reputation and mitigate risks associated with climate-related regulations and consumer preferences. Our primary goal of working with this model is to identify the optimal transition investment ratio that balances financial performance with sustainability objectives, ensuring that companies thrive economically while they are still able to contribute positively to the environment.

In practice, the Transition Investment Ratio offers insights into how effectively a company aligns its financial resources with its sustainability goals. A higher value of this ratio signifies a greater investment in low-carbon initiatives, leading to long-term cost savings and competitive advantages. For instance, IKEA's allocation of over 3 billion pounds toward a renewable energy infrastructure exemplifies the benefits of a strategic adjustment in this ratio (\cite{ikea2020sustainability}). By investing in solar and wind energy projects, IKEA not only reduces its operational reliance on fossil fuels but also achieves improved operational efficiency and lower carbon emissions. Improved operational efficiency can manifest in reduced energy costs, as renewable energy sources often lead to lower utility bills. IKEA’s renewable energy infrastructure allows it to produce a substantial portion of its energy needs in a sustainable manner, ultimately lowering the company's operational costs and enhancing its energy independence. Additionally, the transition to renewable energy directly impacts carbon emissions, as utilizing solar and wind energy decreases IKEA’s greenhouse gas footprint. This shift aligns with regulatory expectations and resonates with consumers who favor environmentally responsible brands. Thus, the Transition Investment Ratio plays a pivotal role in our model as it encapsulates the strategic allocation of resources toward sustainability and driving both operational efficiency and lower carbon emissions. 

As previously mentioned, undergoing a low-carbon transition and investing in this transition can decrease production costs and increase selling prices for the company due to improved operational efficiency, stronger brand value, and an alignment with consumer preferences. In the model, we assume both the production costs and the selling prices to be constant for simplicity. This assumption allows us to isolate the decision-making process regarding the resource allocation and emphasize the role of the Transition Investment Ratio in balancing financial performance and sustainability objectives without introducing additional complexities. 

\textbf{Enterprise Low-carbon Production Efficiency Coefficient ($k$)}.  
This coefficient measures the improvement in the firm's production efficiency resulting from low-carbon initiatives, with a value greater than 1 indicating a gain in efficiency due to cleaner technologies. In our model, we assume that a higher value of $k$ indicates greater production efficiency of the firm after investing in the transition to sustainable practices. For instance, Siemens reported a remarkable 20\% increase in production efficiency after transitioning from traditional fossil fuel-based energy to renewable energy sources and incorporating energy-efficient manufacturing technologies (\cite{siemens2019}). This transition not only reduced their carbon footprint but also streamlined their operations, demonstrating that investments in sustainability can yield great returns in efficiency. 

\textbf{Original Productivity Coefficient ($\beta$)}.
This coefficient is set to range between 0 and 1. A lower value indicates that, given a fixed amount of raw materials, the firm’s production efficiency is relatively low, resulting in fewer products being produced per unit of input, which reflects suboptimal resource utilization. Conversely, a value closer to 1 suggests higher efficiency in material utilization, allowing the enterprise to produce more output with the same raw material input, indicating both optimized resource allocation and production process efficiency. Changes in this coefficient can thus reflect the current production efficiency of the enterprise and offer insights into potential improvements under low-carbon transition strategies. There is some evidence to suggest that many firms that start with below-average efficiency have successfully turned their operations around after investing in cleaner technologies. Notably, research in \cite{weng2015effects} indicates that companies in the regions with lower initial productivity often experience the most substantial gains when transitioning to more efficient, low-carbon systems. By enhancing productivity through sustainable practices, these firms not only are able to improve their operational performance but they also are able to position themselves for greater profitability in the long run. Thus, the firms with lower \(\beta\) values should prioritize upgrades to enhance productivity and better align with sustainability goals. 

\textbf{Carbon Price ($B$)}.
This variable captures the cost of emissions imposed by regulatory frameworks and is essential to a company's profitability. This price is sensitive to regulatory or policy changes, with high carbon prices creating a financial pressure on firms to reduce emissions. When the market carbon price is elevated, it can diminish a company's profit margins, compelling businesses to reassess their operational strategies. In response to these challenges, companies are incentivized to invest in cleaner technologies and adopt sustainable practices to mitigate their carbon liabilities. As firms strive to maintain profitability while adhering to environmental regulations, understanding and managing the implications of carbon pricing becomes crucial for their long-term success.

\textbf{Carbon Intensity ($I$)}. 
This variable refers to the amount of carbon dioxide (CO2) emissions produced per unit of output. It
is typically measured in terms of emissions per product manufactured or per unit of energy consumed. It serves as a critical metric for companies transitioning to low-carbon production, as reducing carbon intensity is essential for minimizing environmental impact while maintaining operational efficiency.

In the literature, carbon intensity is usually defined as a measure of the environmental performance of a company's production processes. For instance, the 
Intergovernmental Panel on Climate Change (IPCC) emphasizes that carbon intensity is a key indicator of how effectively an organization utilizes resources and minimizes emissions in its operations (\cite{change2014mitigation}). Similarly, the World Resources Institute (WRI) defines carbon intensity as a means to assess the relationship between output and emissions, allowing for comparisons across industries and over time (\cite{wri2015}). In addition, carbon intensity is defined more specifically as the ratio of expected annual emissions to expected revenue in \cite{fang2019sustainable}. This formulation highlights the firm's contribution to climate change per unit of revenue, thus linking environmental impact directly to economic output and enabling the pricing of emissions-related risks in financial evaluations.

\textbf{Flow and Interaction between Variables}.
The interaction between these variables allows us to assess the financial, operational, and environmental trade-offs involved in a low-carbon transition for companies. For example, increasing the transition investment ratio ($\alpha$) may initially raise firm's production costs ($c$), but this can lead to greater efficiency ($k$) and reduced carbon intensity ($I$) for the firm over time. Firms in markets with high carbon prices ($B$) are more likely to invest in low-carbon technologies, seeing long-term benefits from both regulatory compliance and enhanced profitability through premium pricing and consumer demand for sustainable products. By considering these variables, companies can strategically navigate the low-carbon transition, balancing short-term financial considerations with long-term environmental and competitive advantages. This paper constructs a highly stylized, quantitative model based on these variables.

The variables of our proposed model are summarized in Table \ref{table1}. 

\begin{table}[H]
    \centering
    \begin{tabular}{|c|c|}
    \hline
         Notation & Description\\
         \hline
         $p$& Selling Price per unit\\
         $c$& Production Cost per unit\\
         $p-c$& Selling Profit per unit\\
         A& Total Asset\\
         $\alpha$& Transition Investment Ratio\\
         $k$& Enterprise Low-carbon Production Efficiency Coefficient\\
         $\beta$& Original Productivity Coefficient\\
         $B$& Carbon Price\\
         $I$& Carbon Intensity\\
         \hline
    \end{tabular}
    \caption{Variables of the Model}
    \label{table1}
\end{table}

\subsection{Optimal Transition Investment Ratio without Government Intervention}

The transition towards a low-carbon economy presents a complex challenge for enterprises, acting as a double-edged sword with respect to both arising opportunities and costs. While embracing low-carbon transformation holds the promise of environmental sustainability and enhanced competitiveness, it also entails a substantial amount of financial investments and operational adjustments for firms. For many enterprises, particularly those operating in carbon-intensive industries, the costs associated with transitioning to cleaner technologies and sustainable practices can be daunting. These costs may include investments in a renewable energy infrastructure, efficiency upgrades, and emission reduction measures, all of which require a substantial amount of capital expenditure and major operational restructuring in order to embark on the transition path.

As a result, some companies may hesitate to fully commit to a low-carbon transformation, fearing the resulting financial burden and potential disruptions to their existing business models. This reluctant proclivity to adopt green practices could undermine a firm's progress towards sustainability goals and exacerbate environmental challenges. In such cases, the role of government intervention becomes relevant in incentivizing and regulating enterprises' low-carbon transformation efforts.

Governments possess the ability to influence corporate behavior through various policy instruments, including subsidies, tax incentives, and regulatory measures. By introducing targeted subsidies or tax breaks for low-carbon investments, governments can help alleviate the financial barriers faced by enterprises and encourage them to pursue sustainable practices. These financial incentives can offset the upfront costs of transitioning to cleaner technologies and provide companies with the necessary resources to invest in green initiatives.

Furthermore, governments can enact regulatory policies that mandate emission reductions, promote energy efficiency, and set emissions targets for industries. By establishing clear and enforceable regulations, governments create a framework that incentivizes companies to prioritize low-carbon transformation and align their business strategies with broader environmental objectives. Regulatory measures can also level the playing field for all companies by imposing costs on carbon emissions and internalizing the externalities associated with pollution, thereby encouraging companies to internalize the true costs of their environmental impact.

Hence, this section aims at analyzing and delineating the optimal low-carbon transition investment strategies for companies, considering both scenarios with and without external government intervention. By comparing these two approaches, we hope to elucidate the efficacy and implications of different pathways toward sustainable transformation within the corporate landscape and to provide valuable insights and suggestions for policymakers, such as government entities, to better support and incentivize effective low-carbon strategies in the corporate sector.



Maximizing profit is a prime objective of a company. This strategic action emanates from the basic principle that profit arises from the difference between total revenue and total costs. Therefore, to effectively navigate and optimize profit margins, it becomes indispensable to dissect the constituents of both total revenue and total costs of the company.
In our model, total revenue is represented by the unit selling price per unit of the product (denoted as $p$) multiplied by the total production volume. The total production volume is represented by
a stylized sigmoid function,  $2 \cdot \text{Sigmoid}(\alpha k) \cdot [(1 - \alpha) \cdot A]^\beta$, where $\operatorname{Sigmoid}(\alpha k)=\frac{1}{1+e^{-\alpha k}}$. This stylized formulation integrates several key variables: the transition investment ratio ($\alpha$), indicating the allocation of total assets towards low-carbon initiatives; the enterprise low-carbon production efficiency coefficient ($k$), denoting the effectiveness of transitioning to cleaner technologies; and the original productivity coefficient ($\beta$), reflecting the baseline production efficiency. The sigmoid function encapsulates the joint impact of transition investments and the firm's production efficiency on production volume, ensuring a realistic representation of the firm's saturation effects. Meanwhile, the term $(1 - \alpha) \cdot A$ captures the portion of the firm's total assets available for production activities, adjusted by the original productivity coefficient $\beta$ to reflect the company's inherent efficiency before investing in the low-carbon transition. Thus, the term \( [(1 - \alpha) \cdot A]^\beta \) represents the firm's total production volume before investing in a low-carbon transition. Simultaneously, the firm allocates a portion \( \alpha \) of its total assets \( A \) towards the low-carbon transition investment. This investment is expected to enhance production efficiency, leading to an overall increase in production volume by a factor of \( 2 \cdot \operatorname{Sigmoid}(\alpha k) \). By combining these various elements, the formula provides a greater understanding of how strategic decisions regarding transition investments and production efficiency influence the overall production capacity of the company.

The design of this production function originates from a Cobb-Douglas production function, which is a widely used model in Economics to link a set of inputs (typically capital and labor) to an output in a production process. Its form with 
constant returns to scale is $Q=A \cdot K^\beta \cdot L^{1-\beta}$, where $Q$ is a total output, $A$ is a scaling or efficiency parameter, $K$ is a capital input, $L$ is a labor input, and $\beta$ is the elasticity of output with respect to capital. In our model, the term $[(1-\alpha) \cdot A]^\beta$ corresponds to $K^\beta$. The sigmoid function, multiplied by 2, acts as an adaptive transformation of the efficiency parameter $A$ in the Cobb-Douglas
production function, reflecting the transition investment's impact on production efficiency. Unlike the traditional Cobb-Douglas function, which considers both capital and labor inputs, our model excludes the labor component due to the specific context of low-carbon investment decisions. This modification is intended to ensure that the production process incorporates diminishing marginal returns to transition investments ( $\alpha$ ) and captures the saturation effects as efficiency reaches its practical limits. Notably, we introduce a scaling factor of `2' at the beginning of the sigmoid function. This adjustment is made primarily to ensure that the sigmoid's baseline value remains at least 0.5 when the input ($\alpha k$) exceeds 0 in our model. By multiplying the sigmoid by `2', we normalize its baseline rate to 1, resulting in a more pronounced increase above 1 as the values of $\alpha$ (Transition Investment Ratio) or $k$ (Enterprise Low-carbon Production Efficiency Coefficient) rise. This modification aligns with our prior expectations about this relationship.

Our results are expected to hold broadly with alternative forms of the production functions exhibitting concavity and diminishing marginal returns, which are the main properties captured by the sigmoid-based function. Once such production function is a Constant Elasticity of Substitution (CES).
The trade-offs between transition investments and production efficiency, as well as the insights on profitability, depend on these shared characteristics and are not restricted to the functional form that we use in our study. However, the CES function, which allows for variable elasticity of substitution between inputs, could confer an additional degree of flexibility and provide further insights into how substitutability between capital and efficiency affects production outcomes. 

Given the above description of the model, we define `Revenue' as:

\begin{equation}
\begin{aligned}
\text { Revenue } & =\text { Selling Price per unit } \times \text { Unit Amount } \\
& =p \times\left\{2 \cdot \text { Sigmoid }(\alpha k) \cdot[(1-\alpha) \cdot A]^\beta\right\} \\
& =p \times\left\{\left(\frac{2}{1+e^{-\alpha k}}\right) \cdot[(1-\alpha) \cdot A]^\beta\right\}
\end{aligned}
\end{equation}

Next, we address the `Cost' component of the model, which comprises both the `Production Cost' and the `Carbon Cost'. Denote `c' as the production cost per unit. The production cost can then be expressed as `c' multiplied by the unit amount (equivalent to the unit amount in the revenue part). Additionally, we incorporate the carbon cost in our model, which is calculated as the product of the `Carbon Intensity', the unit amount, and the `Carbon Price'. Here, the `Carbon Intensity' is defined as $\left(2-\frac{2}{1+e^{-\alpha k}}\right)$, which is a value set to  range between 0 and 1. When $\alpha k$ is zero (indicating no transition investment), the carbon intensity reaches its highest level of 1. Conversely, as transition investment increases, the carbon intensity decreases, indicating a reduction in emissions due to greater transition efforts. As mentioned, we define `Carbon Intensity' as the `Carbon Emission' divided by the total production volume, aligning with the traditional understanding of `Carbon Intensity' discussed earlier. Based on these considerations, we define the `Cost' component in the model as:

\begin{equation}
\begin{aligned}
\text{Cost} = &\text{ Production Cost + Carbon Cost} \\
= &\text{ Production Cost Per Unit} \times \text{Unit} + \underbrace{\text {Carbon Intensity } \times \text { Unit }}_{\text {Carbon Emission }} \times \text{Carbon Price} \\
 =& c \cdot\left\{\left(\frac{2}{1+e^{-\alpha k}}\right) \cdot[(1-\alpha) \cdot A]^\beta\right\} \\
& +\left(2-\frac{2}{1+e^{-\alpha k}}\right) \cdot\left\{\left(\frac{2}{1+e^{-\alpha k}}\right) \cdot[(1-\alpha) \cdot A]^\beta\right\} \cdot B
\end{aligned}
\end{equation}

\noindent Thus, the total profit $\pi$ is calculated as: 

\begin{equation}
\begin{aligned} 
\pi&=\text { Revenue }-\operatorname{Cost} \\ 
& =\left[p-c-\left(\frac{2 e^{-\alpha k}}{1+e^{-\alpha k}}\right) \times B\right] \times\left[\frac{2}{1+e^{-\alpha k}} \times[(1-\alpha) \times A]^\beta\right]
\end{aligned}
\label{equ:pi}
\end{equation}

To determine the optimal value of $\alpha$, we take the first-order derivative of the total profit with respect to $\alpha$ and set the resulting equation to be eqiual to 0. This yields a necessary first-order condition (FOC) from the firm's profit optimization problem:

\begin{equation}
\begin{aligned}
\frac{2((1-\alpha) A)^\beta\left(-\frac{2 B e^{\alpha(-k)}}{e^{\alpha(-k)}+1}-c+p\right)}{e^{\alpha(-k)}+1}=0
\end{aligned}
\end{equation}

\noindent To determine whether the solution to the FOC yields a maximum value for the firm's profit, we obtain a second-order derivative with respect to $\alpha$:

\begin{equation}
\begin{aligned}
& \frac{1}{(\alpha-1)^2\left(e^{\alpha k}+1\right)^3} 2 \alpha e^{\alpha k}(A-\alpha A)^\beta \left(-(\alpha-1)^2 k^2\left(e^{\alpha k}-1\right)\right.\\
& \left.+(\beta-1) \beta\left(e^{\alpha k}+1\right)^2+2(\alpha-1) \beta k\left(e^{\alpha k}+1\right)\right)
\end{aligned}
\end{equation}

\noindent We observe that the value is less than 0 when $\alpha$ and $\beta$ are both within the range of (0, 1). This leads us to conclude that the optimal value of $\alpha$ indeed maximizes the firm's profit in keeping with our prior expectation.

As a first step, we try to understand how changes in specific variables affect the optimal transition investment ratio. For instance, we wish to determine how the optimal value of $\alpha$ adjusts when the carbon price ($B$) increases from 0 to 2. This analysis will elucidate the impact of carbon pricing on the optimal transition investment ratio for the firm. However, explicitly expressing $\alpha$ in terms of other variables analytically is highly intractable. The equation features exponential terms such as \(e^{\alpha(-k)}\), which complicate the isolation of \(\alpha\) since these terms do not simplify easily into a linear or polynomial form. Additionally, the equation includes a combination of linear and non-linear terms involving multiple variables (\(A\), \(B\), \(c\), \(p\), \(\beta\), and \(k\)), further complicating the analysis. The presence of \(\alpha\) in both the numerator and the denominator within the fractional expressions adds another layer of difficulty in the analysis. Therefore, below we will resort to numerical methods to assess how the optimal transition investment ratio of the firm varies in response to incremental parameter changes.


We conduct a numerical analysis of each critical variable in the model, which  affects the optimal transition investment ratio (\(\alpha\)). Recall that the optimal transition investment ratio represents the proportion of firm's financial resources allocated towards its low-carbon initiatives relative to its total investment. A key variable is a specific variable within the model that we systematically vary to observe its impact on the optimal \(\alpha\). By focusing on one variable at a time, we can isolate its effect and gain deeper insights into its role within the model. 

While holding all other variables constant in the model, we determine the optimal value of \(\alpha\) for a range of values of a specific key variable. We then graphically represent the relationship between this key variable and the optimal \(\alpha\). Additionally, we will present the maximum profit achievable for different values of the key variable. This approach allows us to visualize how changes in one variable influence both the optimal investment decision and the profitability of the company.

The baseline values of the variables in the model are provided in Table \ref{table2}. These values are set within a reasonable range for illustrative purposes and can be adjusted by readers interested in exploring scenarios across different industries. Building on this foundation, we will systematically vary one variable at a time to different values and analyze the resultant changes in the value of $\alpha$.

\begin{table}[H]
    \centering
    \begin{tabular}{|c|c|c|}
    \hline
         Variable Name& Description & Baseline Value\\
         \hline
         $p$& Selling Price per unit & 3.6\\
         $c$& Production Cost per unit & 1.6\\
         $p-c$& Selling Profit per unit & 2\\
         A& Total Asset & 100\\
         $k$& Enterprise Low-carbon Production Efficiency Coefficient & 2\\
         $\beta$& Original Productivity Coefficient & 0.8\\
         $B$& Carbon Price & 1\\
         \hline
    \end{tabular}
    \caption{Baseline Values of Variables in the Model}
    \label{table2}
\end{table}

\begin{figure}[H]
\includegraphics[width=13cm]{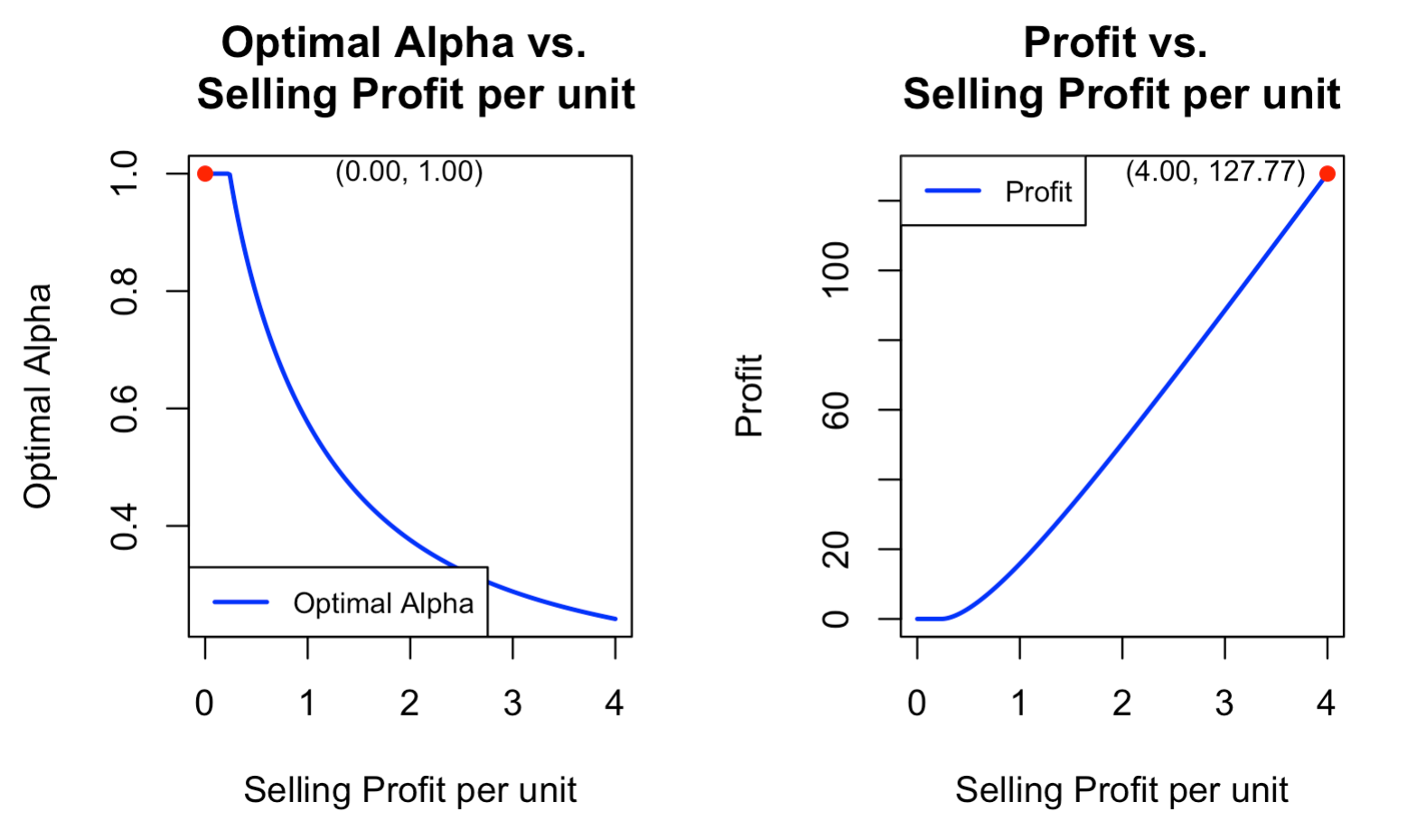}
\centering
\caption{Optimal Level of Alpha and Profit Graph w.r.t. $p-c$}
\label{figpc}
\end{figure}

\begin{figure}[H]
\includegraphics[width=12cm]{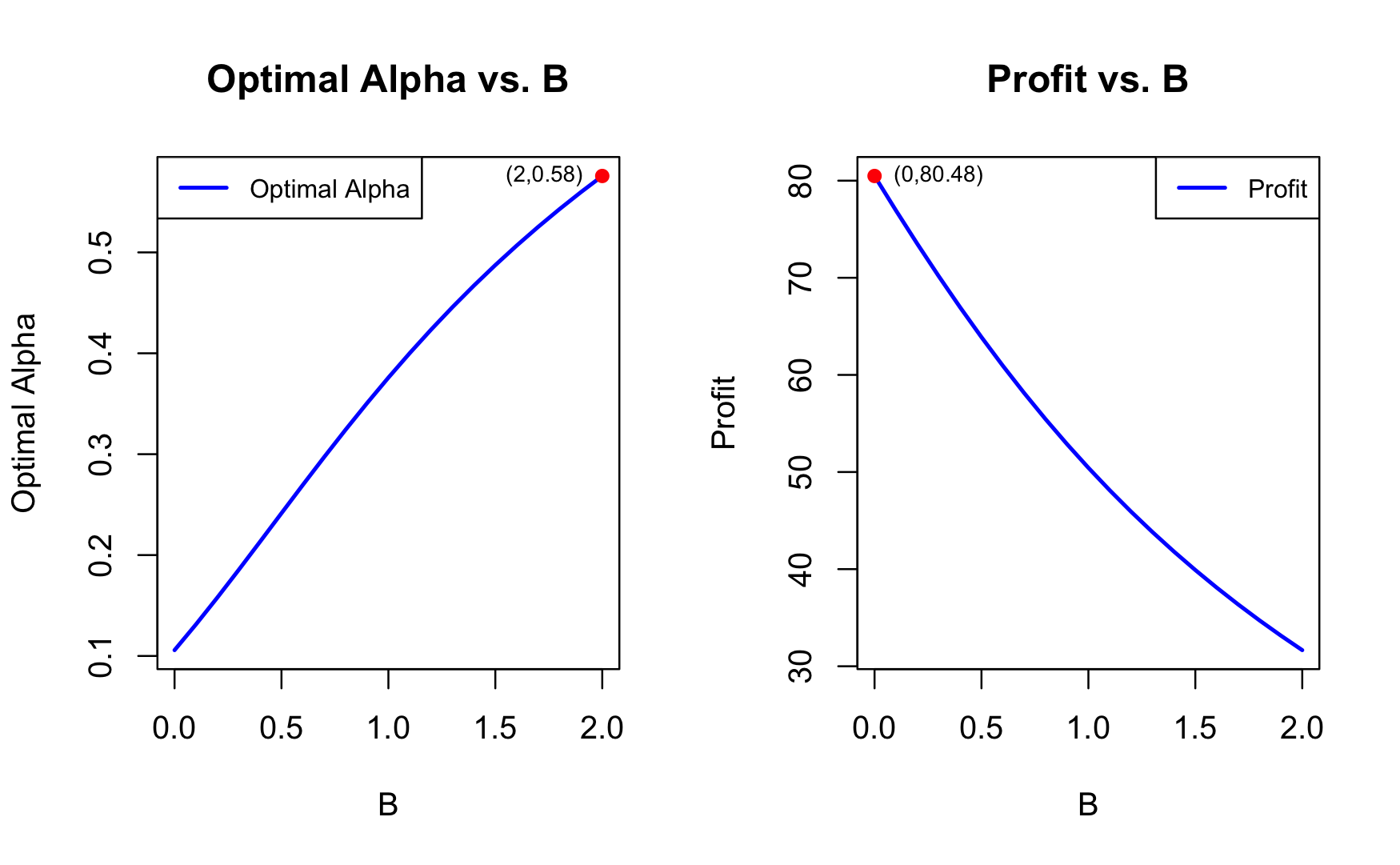}
\centering
\caption{Optimal Level of Alpha and Profit Graph w.r.t. $B$}
\label{figB}
\end{figure}

\begin{figure}[H]
\includegraphics[width=12cm]{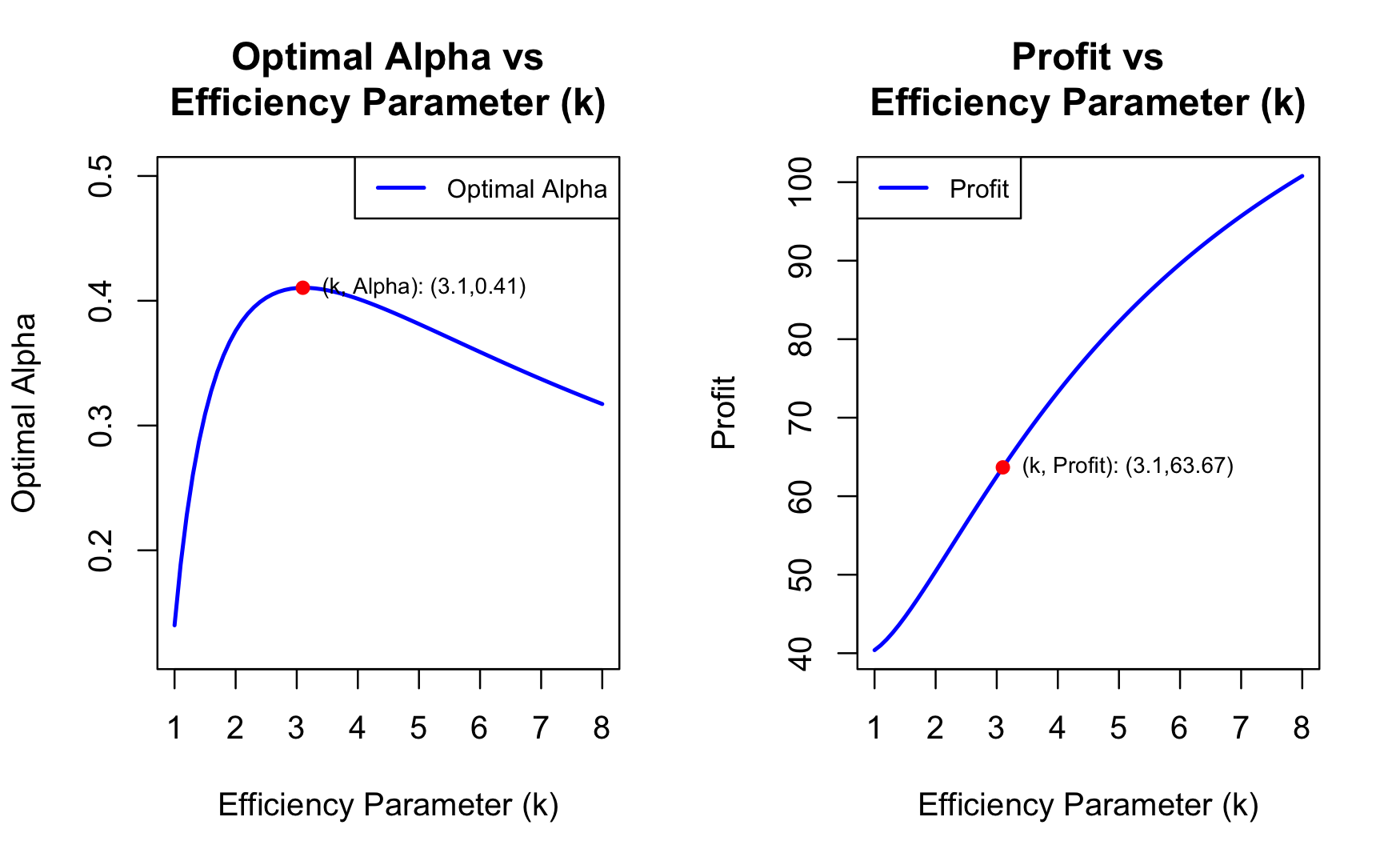}
\centering
\caption{Optimal Level of Alpha and Profit Graph w.r.t. $k$}
\label{figk}
\end{figure}

\begin{figure}[H]
\includegraphics[width=12cm]{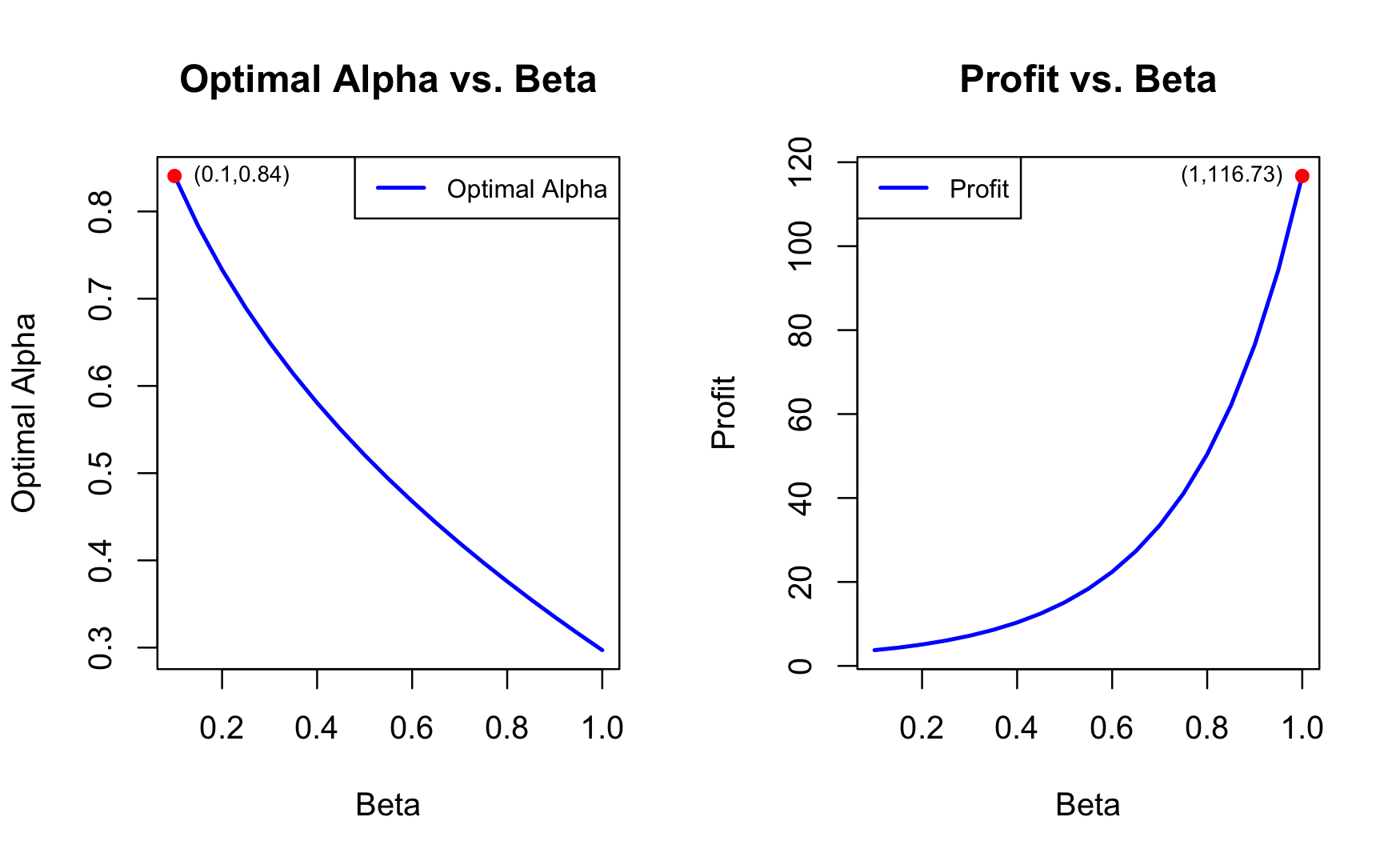}
\centering
\caption{Optimal Level of Alpha and Profit Graph w.r.t. $\beta$}
\label{figbeta}
\end{figure}

From Figure \ref{figpc} to Figure \ref{figbeta}, the left plot illustrates how changes in the key variable in the model affect the optimal transition investment ratio, with the goal of maximizing profit under each scenario. The right plot shows the corresponding maximum profit achieved under the optimal transition investment as the key variable changes. We will explain each figure in detail below.

\noindent \textbf{Relationship Between Profit Margin (\(p-c\)) and Optimal Alpha} (Figure \ref{figpc}). Figure \ref{figpc} reveals a clear pattern where the optimal value of $\alpha$ for the firm remains high and constant initially, even when the selling profit per unit is very low or negative, suggesting a strategic focus on substantial investments in machinery upgrades or efficiency improvements during periods of low profitability. This high investment level, indicated by a stable high value of $\alpha$, which is likely to reflect a counterintuitive but proactive approach: companies invest heavily during financially challenging times to enhance production efficiency and reduce operational costs, aiming at setting the groundwork for future competitiveness and sustainability.

As the selling profit per unit begins to increase for the firm, the optimal value of $\alpha$ sharply declines, indicating that further heavy investments in transformation by the firm are no longer necessary. The decrease in the value of $\alpha$ suggests that the earlier investments by the firm have begun to yield improved operational efficiencies and higher profits, reducing the need for a continued high expenditure in transformations. This strategic shift emphasizes a dynamic financial strategy where initial heavy investments by the firm during low-profit periods are scaled back as profitability improves, aligning the investment levels of the firm with the diminishing marginal benefits of further transformations.

\noindent \textbf{Effect of Increasing Carbon Prices (\(B\)) on Optimal Alpha} (Figure \ref{figB}). As shown in Figure \ref{figB}, when the carbon pricing variable \( B \) increases, it raises operational costs for the firm due to higher carbon taxes. In response, the optimal investment ratio \( \alpha \) also increases, as shown in the left plot. This adjustment indicates a strategic move by the firm to allocate more resources towards low-carbon transition initiatives, which enhances production efficiency and reduces carbon emissions, thereby helping to offset the financial burden imposed by the carbon tax.

However, the right plot shows that despite these adjustments, a higher carbon tax still leads to a reduction in overall profit. Even with an optimal increase in the low-carbon investment ratio \( \alpha \), the profit trend declines as \( B \) increases. This underscores the financial impact of rising carbon costs on the firm. Nonetheless, by investing strategically in low-carbon transition, the firm can mitigate the extent of this profit decrease, cushioning some of the adverse effects of carbon pricing. In essence, while higher carbon taxes inevitably compress profits, allocating resources towards reducing the carbon footprint can help alleviate the downward pressure on profitability, offering a partial buffer against the increasing tax burden.

\noindent \textbf{Impact of Efficiency Parameter (\(k\)) on Optimal Alpha and Profit} (Figure \ref{figk}). Combining insights from both plots in Figure \ref{figk}, we can observe that as the efficiency parameter \( k \) increases, profit consistently rises, while the optimal investment ratio \( \alpha \) peaks at around \( k \approx 3.1 \) and then gradually declines. This suggests that although higher \( k \) values are beneficial for the company’s profitability, it does not require continuously increasing the investment ratio in low-carbon initiatives. Initially, when \( k \) is low, the marginal benefit from efficiency gains is high, encouraging a higher allocation to low-carbon investments. However, as \( k \) grows, the marginal efficiency gains diminish, making it less necessary to maintain a high \( \alpha \). At higher \( k \) values, profits continue to increase due to the enhanced production efficiency achieved by prior investments, allowing the company to reduce the investment ratio \( \alpha \) while still benefiting from increased profits.

\noindent \textbf{Influence of Original Productivity Coefficient (\(\beta\)) on Optimal Alpha and Profit} (Figure \ref{figbeta}). The plots in Figure \ref{figbeta} demonstrate that as the Original Productivity Coefficient (\(\beta\)) increases, profit grows exponentially while the optimal transition investment ratio (\(\alpha\)) declines. A higher \(\beta\) indicates strong baseline productivity, reducing the need for extensive low-carbon investments since the firm’s production efficiency is already high. This makes the additional benefits from low-carbon investments less pronounced, allowing the firm to allocate fewer resources to \(\alpha\). The exponential growth in profit with rising \(\beta\) highlights the impact of inherent productivity on profitability, as firms with high baseline efficiency can achieve substantial profits without heavy investment in low-carbon initiatives. The decrease in optimal \(\alpha\) reflects diminishing marginal returns on low-carbon investments for high-\(\beta\) firms, where the benefits of further investment in \(\alpha\) are minimal.

The sensitivity analysis is conducted to evaluate the impact of variations in critical variables on the optimal $\alpha$, the maximum objective value, and the profit change rate. The variables scrutinized include the enterprise low-carbon production efficiency coefficient (\(k\)), production cost per unit (\(c\)), original productivity coefficient (\(\beta\)), and carbon price per unit (\(B\)). Adjustments for each parameter are performed at ±10\% and ±50\% based on the baseline values in Table \ref{table2}, except for \(\beta\), where the upper limit is constrained to 1.00 due to inherent limitations. The results are presented in the Table \ref{tab:sensitivity_analysis} and the analysis of each variable and parameter is reported below. 

\begin{table}[H]
\centering
\begin{tabular}{|c|c|c|c|c|c|c|}
\hline
\multicolumn{6}{|c|}{$k$ (Enterprise Low-carbon Production Efficiency Coefficient)} \\
\hline
Parameter Value & 1.0 & 1.8 & 2.0 & 2.2 & 3.0 \\
\hline
Optimal Alpha & 0.1399250 & 0.3560002 & 0.3758660 & 0.3896782 & 0.4102122 \\
\hline
Max Objective Value & 40.38998 & 48.03570 & 50.43508 & 52.88084 & 62.52014 \\
\hline
Alpha Change Rate (\%) & -62.77263 & -5.285352 & 0.000000 & 3.674763 & 9.137887 \\
\hline
Profit Change Rate (\%) & -19.916887 & -4.757367 & 0.000000 & 4.849314 & 23.961613 \\
\hline
\multicolumn{6}{|c|}{$c$ (Production Cost Per Unit)} \\
\hline
Parameter Value & 0.80 & 1.44 & 1.60 & 1.76 & 2.40 \\
\hline
Optimal Alpha & 0.3011413 & 0.3571930 & 0.3758660 & 0.3970913 & 0.5191683 \\
\hline
Max Objective Value & 80.82201 & 56.40787 & 50.43508 & 44.53826 & 22.16245 \\
\hline
Alpha Change Rate (\%) & -19.880685 & -4.967987 & 0.000000 & 5.647027 & 38.125899 \\
\hline
Profit Change Rate (\%) & 60.249582 & 11.842530 & 0.000000 & -11.69191 & -56.05747 \\
\hline
\multicolumn{6}{|c|}{$\beta$ (Original Productivity Coefficient)} \\
\hline
Parameter Value & 0.40 & 0.72 & 0.80 & 0.88 & 1.00 \\
\hline
Optimal Alpha & 0.5806144 & 0.4107206 & 0.3758660 & 0.3430713 & 0.2971025 \\
\hline
Max Objective Value & 10.33973 & 36.31521 & 50.43508 & 70.34946 & 116.72607 \\
\hline
Alpha Change Rate (\%) & 54.473780 & 9.273146 & 0.000000 & -8.725114 & -20.955217 \\
\hline
Profit Change Rate (\%) & -79.49894 & -27.99613 & 0.000000 & 39.48516 & 131.43825 \\
\hline
\multicolumn{6}{|c|}{$B$ (Carbon Price Per Unit)} \\
\hline
Parameter Value & 0.5 & 0.9 & 1.0 & 1.1 & 1.5 \\
\hline
Optimal Alpha & 0.2414888 & 0.3505336 & 0.3758660 & 0.4002088 & 0.4874011 \\
\hline
Max Objective Value & 63.88578 & 52.87325 & 50.43508 & 48.11505 & 39.90827 \\
\hline
Alpha Change Rate (\%) & -35.751365 & -6.739751 & 0.000000 & 6.476445 & 29.674174 \\
\hline
Profit Change Rate (\%) & 26.669326 & 4.834262 & 0.000000 & -4.600030 & -20.871997 \\
\hline

\end{tabular}
\caption{Sensitivity Analysis Results}
\label{tab:sensitivity_analysis}
\end{table}

\noindent \textbf{Enterprise Low-Carbon Production Efficiency Coefficient $(k)$ }.
The sensitivity analysis for $k$ reveals that as $k$ increases from 1.0 to 3.0, the optimal $\alpha$ rises from 0.14 to 0.41, which indicates that improving a firm's low-carbon production efficiency encourages a larger investment by the firm in low-carbon technologies. This trend is further reflected in the maximum objective value, which grows from 40.39 to 62.52, showing that better firm's efficiency leads to higher profitability for the firm. These results highlight a strong connection between firm's low-carbon production efficiency and its profitability.

Moreover, the $\alpha$ change rate starts with a large negative value (-62.77\%) when $k=1.0$ and makes a transition to a moderately positive change of $9.14 \%$ by $k=3.0$. Similarly, the firm's profit change rate shifts from a notable loss ($-19.92\%$) at $k=1.0$ to a relatively large gain $(23.96\%$) at $k=3.0$. This indicates that as efficiency improves, firms not only increase their levels of investment in cleaner technologies but also experience greater profitability. These results underline the importance of improving low-carbon production efficiency, as it can serve as a key driver for both firm's optimal investment and its profit maximization.

\noindent \textbf{Production Cost Per Unit (c)}.
As $c$ increases from 0.80 to 2.40, the optimal $\alpha$ rises from 0.30 to 0.52, suggesting that firms must invest more to counterbalance higher production costs. However, this comes at the cost of its profitability, as the maximum objective value drops sharply from 80.82 to 22.16, showing that rising production costs significantly undermine a company's profitability potential.

The $\alpha$ change rate starts at $-19.88\%$ for $c=0.80$, meaning that lower production costs lead to less investment in low-carbon technology. As $c$ rises, the $\alpha$ change rate becomes positive (38.13\%), showing that companies need to increase their investments to cope with higher costs. Profit change rates, meanwhile, vary drastically, reaching a peak at $60.25\%$ for $c=0.80$, followed by a steep decline to $-56.06\%$ at $c=2.40$. This result illustrates the sensitivity of firm's profitability to changes in its production costs and highlights the point that controlling production costs is crucial for firms whose goal is to sustain profits while investing in low-carbon solutions. If its production costs rise unchecked, the firm may struggle to maintain its profitability, even with optimal investments.

\noindent \textbf{Original Productivity Coefficient $(\beta)$ }. 
When it comes to the value of $\beta$, the sensitivity analysis overall shows a unique pattern: as $\beta$ increases from 0.40 to 1.00, the optimal $\alpha$ decreases from 0.58 to 0.30. This indicates that as its productivity improves, the firm requires less investment in low-carbon technologies to maintain profitability. Simultaneously, the maximum objective value surges from 10.34 to 116.73, highlighting the substantial impact of productivity on overall profitability of the firm. This result emphasizes the fact that improving productivity not only reduces the required investment by the firm but also dramatically boosts the firm's profit potential.

The change rate of $\alpha$ starts at a high level (54.47\%) when $\beta=0.40$ and drops to $-20.96 \%$ at $\beta=1.00$, showing that better productivity substantially reduces the need for the firm to increase the level of investment. In terms of profitability, the profit change rate for the firm ranges widely, from $-79.50\%$ to a remarkable $131.44\%$. This further illustrates how vital firm's productivity improvements are for enhancing its profitability. Essentially, better productivity allows firms to generate substantially higher profits with a reduced level of investment, making this an essential factor in optimizing their financial strategies of the firm.

\noindent \textbf{Carbon Price Per Unit $(B)$}.
Finally, the analysis of $B$ shows that as carbon prices increase from 0.50 to 1.50, the optimal $\alpha$ rises from 0.24 to 0.49, suggesting that the firm needs to allocate more resources to low-carbon production as carbon prices rise. However, this comes at a cost, as the maximum objective value declines from 63.89 to 39.91, showing that higher carbon prices can lead to a reduced level of profitability.

The change rate of $\alpha$ is consistently negative, starting at $-35.75\%$ for $B=0.50$ and improving slightly to $-6.74\%$ at $B=1.10$, indicating that while increased carbon pricing necessitates the firm to invest more, its negative impact on profitability is relatively moderate. The change rate of the profit peaks at $26.67\%$ for $B=0.50$ and drops to $-20.87\%$ at $B=1.50$. This suggests that while carbon pricing does affect firm's profitability, the sensitivity is not as severe for the firm compared to other variables. Companies can mitigate the impact of higher carbon prices through improved efficiency and strategic investments in cleaner technologies, ensuring that their profitability does not drastically diminish as carbon prices rise.

In comparing the variables in the model, it is clear that the original productivity coefficient $(\beta)$ is by far the most sensitive. Changes in the value of $\beta$ lead to substantial shifts in the optimal value of $\alpha$, the maximum objective value, and the change rate of the profit. The strong relationship between productivity and profitability suggests that firms should prioritize improving productivity to maximize financial outcomes.

Conversely, the carbon price per unit $(B)$ is the least sensitive variable in the framework. While changes in $B$ do affect both the optimal $\alpha$ and the profitability, the overall impact for the firm is moderate compared to the other factors. Firms can manage the effects of carbon pricing through strategic investments in cleaner technologies, making it a more controllable factor in its financial performance.

In conclusion, firms should focus heavily on productivity improvements $(\beta)$ to maximize their profitability while carefully managing their production costs $(c)$ and their low-carbon production efficiency ($k$) to optimize investments. Carbon pricing $(B)$, while important, presents a more manageable challenge to firms relative to the other factors.


Initially, we have focused on a one-period model to evaluate the impact of transition investment on enterprise profitability. This static model provides basic but useful insights into the immediate effects of various variables on a company's profit. However, real-world scenarios are dynamic in nature and evolve over time. Therefore, as a next logical step, we extend this one-period model to a multi-period setting to capture the evolving nature of the variables and some of the parameters in the model and their long-term impacts.

To achieve this goal, we revise the profit function to account for time dependency, reflecting the changes in key variables in the model over multiple periods. The time-varying profit function is defined as follows:

\begin{equation}
\begin{aligned}
\pi_t=\left[p-c_t-\left(\frac{2 e^{-\alpha k_t}}{1+e^{-\alpha k_t}}\right) \times B_t\right] \times\left[\frac{2}{1+e^{-\alpha k_t}} \times\left[(1-\alpha) \cdot A_t\right]^{\beta_t}\right]
\end{aligned}
\end{equation}

This formulation captures the dynamic nature of transition investments, where $p$ is the selling price per unit (assume that $p$ is a constant), $c_t$ is the production cost per unit at time $t, \alpha$ is the transition investment ratio, $k_t$ is the enterprise low-carbon production efficiency coefficient at time $t, B_t$ is the carbon price at time $t$, and $\beta_t$ is the original productivity coefficient at time $t$. By incorporating these time-varying variables, we can more accurately model the impact of transition investments over multiple periods.

The resulting multi-period objective function can then be expressed as:
\begin{equation}
\begin{aligned}
\max \left[\pi_t+g \pi_{t+1}+g^2 \pi_{t+2}+\ldots+g^{n-1} \pi_{t+n}\right]
\end{aligned}   
\end{equation}
where $g$ is a discount factor. This objective function accounts for firm's profits in the current period $\left(\pi_t\right)$ and its discounted future profits ($\pi_{t+1}, \pi_{t+2}$, and so on), providing a more complete view of the impact of transition investments for the firm. The inclusion of future profits, discounted to the present time by the factor $g$, acknowledges the time value of money, opportunity costs, and inherent risks faced by the company. The discounting factor also acknowledges the uncertainties associated with long-term projections, ensuring a more realistic assessment of the enterprise's financial trajectory.

By using this multi-period framework, we wish to determine the optimal value of $\alpha$ that maximizes the overall profit of the firm across multiple periods. The multi-period model is particularly useful for understanding the long-term implications of transition investments for the firm. Transitional investments typically incur immediate costs but generate substantial future benefits for the firm. For example, investments in low-carbon technologies might initially increase production costs for the firm due to the capital expenditure required for new equipment or processes. However, over time, these investments can lead to substantial efficiency gains, cost reductions, and improved regulatory compliance, which enhance future profits of the company.

\noindent \textbf{Short-Term vs. Long-Term Observation Periods.}
Our goal is to find the optimal value of $\alpha$ to achieve the maximum objective function value for the firm over different observation periods. For simplicity, we will compare a shorter-term period (of 3 years) with a longer-term period (of 6 years).
The Short-Term Period allows us to understand the immediate and near-future impacts of transition investments for the firm. It is suitable for enterprises looking for quick wins and immediate adjustments in their strategies.
The Long-Term Period captures the long-term benefits and potential drawbacks of transition investments faced by the company. It helps the enterprises' plan strategically for sustained growth and ensures compliance with evolving regulations and market conditions
over time.

By comparing these two periods, we can evaluate how the optimal value of $\alpha$ varies based on the length of the planning horizon for the company. This comparison will provide important insights into the trade-offs between short-term gains and long-term sustainability, guiding enterprises in making informed decisions about their transitional investments. The numerical exercise will be presented in the next subsection to illustrate these findings.

\subsection{Decarbonization Scenario Analysis}


To illustrate the practical application of our multi-period model, we examine four decarbonization scenarios by means of a numerical method: Immediate, Quick, Slow, and No Decarbonization. Decarbonization scenarios are strategic frameworks used to explore and analyze the pathways by which industries, economies, and societies can make a transition from high-carbon to low-carbon or carbon-neutral operations. These scenarios are designed to address the urgent need to mitigate climate change by reducing GHG emissions, particularly carbon dioxide (CO2), which is viewed as a major contributor to human-induced global warming. Each decarbonization scenario represents a different approach, pace, and scale of implementing low-carbon technologies, policies, and practices.

These scenarios draw inspiration from the pathways outlined in IPCC’s Representative Concentration Pathways (RCPs) and Shared Socioeconomic Pathways (SSPs) (\cite{ipcc_sr15, ipcc_ar6}), which provide a framework for modeling emissions trajectories under varying socio-economic, policy, and technological conditions. For example, the Immediate Decarbonization scenario aligns with the stringent measures seen in RCP1.9/SSP1-1.9, where immediate, large-scale investments aim at limiting a global temperature rise to 1.5°C (\cite{ipcc_sr15}). Conversely, the No Decarbonization scenario parallels RCP8.5/SSP5-8.5, representing a "business-as-usual" trajectory with minimal climate action and escalating carbon emissions (\cite{ipcc_ar6_syr}).

By analyzing these different scenarios, we can better understand how decarbonization strategies impact enterprise profitability over multiple periods and identify the optimal transition investment ratio ($\alpha$) for each strategy.

\noindent \textbf{Immediate Decarbonization.}
In this scenario, significant and immediate investments are made in low-carbon technologies and practices. The goal is to achieve rapid reductions in carbon emissions as quickly as possible. This approach mirrors the aggressive pathways described in RCP1.9/SSP1-1.9, where rapid decarbonization is pursued through stringent regulatory measures, substantial financial incentives, and rapid deployment of renewable energy and energy efficiency technologies (\cite{ipcc_sr15}). Immediate decarbonization is typically driven by strong political will and societal urgency to address climate change impacts swiftly. While effective in achieving fast emissions reductions, it may entail higher short-term costs and risks of over-investment.

\noindent\textbf{Quick Decarbonization.}
Quick Decarbonization involves a fast-paced yet balanced transition. Investments in low-carbon technologies are substantial but spread out over a few years rather than all at once. This scenario resembles the pathways described in RCP2.6/SSP1-2.6, where emissions reductions are significant but distributed over time to balance economic feasibility and technical constraints (\cite{ipcc_ar6, iea_wes2022}). Quick decarbonization includes strategies such as accelerated deployment of renewable energy, electrification of transportation, and enhanced energy efficiency measures. It offers a cost-effective and pragmatic approach to achieving climate goals while reducing the economic and logistical burden of an immediate overhaul.

\noindent \textbf{Slow Decarbonization.}
The Slow Decarbonization scenario adopts a more gradual approach to transitioning to a low-carbon economy. Investments and regulatory changes are implemented over a longer period of time, allowing for a smoother transition with less immediate economic disruption. This approach corresponds to RCP4.5/SSP2-4.5, a “middle-of-the-road” pathway where moderate policies and socio-economic factors lead to incremental emissions reductions over time (\cite{ipcc_ar6_syr}). While less disruptive, this scenario risks delaying critical action, potentially increasing long-term mitigation costs and reliance on future technological breakthroughs.

\noindent\textbf{No Decarbonization.}
In the No Decarbonization scenario, no substantial efforts are made to reduce carbon emissions. This approach parallels the "business-as-usual" trajectory represented by RCP8.5/SSP5-8.5, where industries and societies continue with their current high-carbon practices (\cite{ipcc_sr15}). Continued reliance on fossil fuels results in escalating carbon emissions, environmental degradation, and increased climate risks (\cite{iea_wes2022, ipcc_ar6_syr}). This scenario is widely regarded as unsustainable, as it amplifies the adverse economic, environmental, and social impacts of climate change over time.


To effectively analyze the aforementioned impacts of different decarbonization strategies, it is essential to understand how key variables in our model evolve under each scenario and the rationale behind these changes. This part delves into the specific modifications of variables such as the enterprise low-carbon production efficiency coefficient $\left(k_t\right)$, production cost per unit ($c_t$), original productivity coefficient ($\left.\beta_t\right)$, and carbon price $\left(B_t\right)$ for each of the decarbonization scenarios: Immediate, Quick, Slow, and No Decarbonization. By doing so, we can better comprehend the short and long-term effects of the transition on firm's profitability and identify the optimal transition investment ratio $(\alpha)$ for the firm in each case.

\noindent\textbf{Immediate Decarbonization.}
In the Immediate Decarbonization scenario, enterprises make substantial investments in low-carbon technologies immediately. This aggressive approach aims at achieving rapid decarbonization and substantial reductions in carbon emissions in the shortest possible time. As a result, the enterprise's low-carbon production efficiency coefficient $\left(k_t\right)$ starts at a very high level due to a substantial initial level of investment, reflecting major improvements in production efficiency of the firm. Over time, as the immediate benefits are realized, $k_t$ decreases gradually. The production cost per unit $\left(c_t\right)$ of the firm decreases quickly as economies of scale and technological efficiencies are realized from the new low-carbon technologies. The original productivity coefficient $\left(\beta_t\right)$ increases initially due to enhanced production capabilities of the firm but stabilizes over time as the effects of initial investments are more fully integrated. The carbon price ($B_t$) remains at a very high level due to stricter regulations and market adjustments that favor low-carbon outputs.

\noindent \textbf{Quick Decarbonization.}
The Quick Decarbonization scenario involves rapid but slightly less aggressive investments compared to the Immediate scenario. Enterprises aim at achieving substantial decarbonization within a relatively short time frame but spread their investments over a few periods of time to balance costs and benefits. Consequently, the enterprise's low-carbon production efficiency coefficient $\left(k_t\right)$ increases quickly but not as sharply as in the case of the Immediate scenario. The efficiency gains are substantial for the firm but spread over a slightly longer period of time. The production cost per unit $\left(c_t\right)$ decreases rapidly but not as dramatically, reflecting more moderate cost reductions. The original productivity coefficient $\left(\beta_t\right)$ shows a steady increase due to technological improvements and better resource allocation by the firm. The carbon price $\left(B_t\right)$ keeps at a relatively high level, driven by increasing regulatory pressures and market responses.

\noindent \textbf{Slow Decarbonization.}
In the Slow Decarbonization scenario, enterprises adopt a gradual approach to transitioning to low-carbon technologies. This approach spreads investments over a longer period of time, allowing enterprises to manage costs and mitigate risks while steadily reducing carbon emissions. As such, the enterprise's low-carbon production efficiency coefficient $\left(k_t\right)$ increases slowly, reflecting incremental improvements in production efficiency for the firm as investments are made gradually. The production cost per unit $\left(c_t\right)$ decreases steadily but slowly, indicating progressive cost reductions over time. The original productivity coefficient $\left(\beta_t\right)$ shows a gradual increase, correlating with incremental enhancements in productivity as new technologies are adopted by the firm over time. The carbon price $\left(B_t\right)$ is set as a relatively low level, aligning with gradual regulatory changes and market adjustments.

\noindent \textbf{No Decarbonization.}
The No Decarbonization scenario assumes that enterprises do not make any major investments in low-carbon technologies. Instead, they continue with their current practices, resulting in minimal changes to carbon emissions and production efficiency. Consequently, there is no need to find the optimal transitional investment ratio under this scenario, as the transitional investment ratio, \(\alpha\), should always be zero.


By analyzing these different scenarios numerically, we can observe how different levels of investment and urgency in transitioning to low-carbon technologies affect key variables of the framework over time. This analysis helps us understand the trade-offs between immediate costs and long-term benefits, guiding enterprises in making informed decisions about their decarbonization strategies. The goal is to determine the optimal value of $\alpha$ that maximizes the overall profit across multiple periods, ensuring both short-term (3-year observation period) profitability and long-term (6-year observation period) sustainability. We will not consider the scenario of `No Decarbonization' in this part, as there is no need to determine the optimal transitional investment ratio \(\alpha\) under this scenario.

\begin{table}[h]
\centering
\begin{tabular}{|c|c|c|c|c|c|c|}
\hline
\textbf{Year} & \textbf{1} & \textbf{2} & \textbf{3} & \textbf{4} & \textbf{5} & \textbf{6} \\
\hline
\multicolumn{7}{|c|}{$k$ (Enterprise Low-carbon Production Efficiency Coefficient)} \\
\hline
Immediate & 5 & 4.5 & 4 & 3.5 & 3 & 2.5 \\
\hline
Quick & 1.5 & 5 & 4.5 & 4 & 3.5 & 3 \\
\hline
Slow & 2 & 3 & 4 & 5 & 4.5 & 4 \\
\hline
\multicolumn{7}{|c|}{$c$ (Production Cost Per Unit)} \\
\hline
Immediate & 1.6 & 1.7 & 1.8 & 2 & 2.2 & 2.4 \\
\hline
Quick & 2.4 & 1.6 & 1.7 & 1.8 & 2 & 2.2 \\
\hline
Slow & 2.2 & 1.9 & 1.6 & 1.7 & 1.8 & 2 \\
\hline
\multicolumn{7}{|c|}{$\beta$ (Original Productivity Coefficient)} \\
\hline
Immediate & 0.9 & 0.85 & 0.8 & 0.75 & 0.72 & 0.7 \\
\hline
Quick & 0.6 & 0.9 & 0.85 & 0.8 & 0.75 & 0.72 \\
\hline
Slow & 0.7 & 0.8 & 0.9 & 0.85 & 0.8 & 0.75 \\
\hline
\multicolumn{7}{|c|}{$B$ (Carbon Price Per Unit)} \\
\hline
Immediate & 4 & 4 & 4 & 4 & 4 & 4 \\
\hline
Quick & 3 & 3 & 3 & 3 & 3 & 3 \\
\hline
Slow & 2 & 2 & 2 & 2 & 2 & 2 \\
\hline
\end{tabular}
\caption{Results for Scenario Analysis}
\label{tab:scenario_analysis}
\end{table}

Table \ref{tab:scenario_analysis} presents various variables under different scenarios of low-carbon transition for enterprises, focusing on production efficiency, costs, productivity, and carbon pricing. 

In the Immediate Transition scenario, enterprises rapidly adopt low-carbon technologies, leading to a substantial increase in firm's production efficiency initially. This efficiency gain diminishes over time due to the fast adaptation rate and potential initial over-investment by the firm, as reflected in the decreasing values of the enterprise's low-carbon production efficiency coefficient \( k \). Production costs per unit rise initially due to high investments in new technologies, reaching their peak in later years as the benefits of the new technologies become more apparent and grounded. The original productivity coefficient \( \beta \) sees a rapid decrease due to the initial disruption and learning curve of the firm associated with adopting new technologies, eventually stabilizing at a lower level over time. High carbon pricing (\( B \)) is a crucial factor in this scenario, as it works to incentivize the rapid adoption of low-carbon technologies to mitigate the associated costs.

In the Quick Transition scenario, enterprises adopt low-carbon technologies at a moderately fast pace. This approach aims at achieving a rapid improvement in production efficiency, though at a slightly slower rate than in the Immediate scenario. As a result, the initial efficiency gains are significant but not as drastic, allowing the enterprise to avoid potential issues related to over-investment. Given the initially low value of \( k \) (starting at 1.5), there is an urgent need to enhance production efficiency quickly. To address this, the enterprise makes a transition investment before the second year, resulting in a sharp increase in \( k \) to 5 by the second year. This value then gradually declines each subsequent year, reflecting the natural efficiency losses due to wear and tear. Production costs per unit begin at a relatively high level (2.4), creating a strong need for transition investment to reduce costs. With the Quick Decarbonization approach, \( c \) drops markedly to 1.6 in the second year. However, it then rises slightly in the following years, likely due to normal depreciation and the ongoing costs associated with maintaining the new technology. Similarly, the original productivity coefficient \( \beta \) starts at a lower level (0.6), emphasizing the necessity for improvement. Due to the transition investment, \( \beta \) reaches a peak of 0.9 in the second year, reflecting increased productivity. Over time, it gradually decreases, aligning with the adaptation phase and the natural reduction in productivity gains. Carbon pricing \( B \) is maintained at a high but stable level (3) in the Quick Decarbonization scenario. This provides a continuous incentive for a firm to adopt low-carbon technologies, without imposing excessive financial burdens on the firm. 

In the Slow Transition scenario, enterprises adopt low-carbon technologies at a gradual pace, resulting in a steady increase in production efficiency without abrupt changes. The Enterprise Low-carbon Production Efficiency Coefficient (\( k \)) begins at a modest value of 2 and gradually rises to 5 by the fourth year before leveling off slightly, indicating a more controlled and gradual improvement. This slower adoption rate minimizes the impact of efficiency losses due to wear and tear over time. The Production Cost Per Unit (\( c \)) starts relatively high at 2.2 but decreases progressively as the benefits of low-carbon technology adoption become more apparent, reaching a stable level at 1.8 by the final year. Such a pattern reflects slower capital outlays and reduced urgency in operational adjustments, leading to consistent long-term savings. The Original Productivity Coefficient (\( \beta \)) also starts at a low value of 0.7 and improves steadily to a peak of 0.9 in the third year, maintaining stability thereafter. This steady incline suggests a smoother, less disruptive learning curve associated with gradual integration of new technologies. Finally, the Carbon Price Per Unit (\( B \)) remains low at 2 throughout the period, applying gentle pressure on firms to adopt low-carbon technologies while allowing them to adjust at a manageable pace, balancing the transition costs with sustained incentives.


The simulation results in Table \ref{tab:scenario_result} reveal an intriguing insight into the relationship between decarbonization strategies and firm profitability over different time horizons. In the short term, the Immediate decarbonization scenario yields the highest profits for the firm, while in the long term, profits are maximized under the Slow decarbonization strategy. The economic intuition behind this pattern lies in the interplay between the upfront investment requirements, efficiency gains, and financial planning associated under the different decarbonization strategies.

\begin{table}[H]
\centering
\begin{tabular}{|l|c|c|c|c|}
\hline
& \multicolumn{2}{|c|}{\textbf{Short-Period}} & \multicolumn{2}{|c|}{\textbf{Long-Period}} \\
\hline
& Optimal $\alpha$ & Max Profit & Optimal $\alpha$ & Max Profit \\
\hline
Immediate & 0.5692 & 173.0291 & 0.6216 & 212.0262 \\
\hline
Quick & 0.541 & 158.3299 & 0.5678 & 235.1407 \\
\hline
Slow & 0.4768 & 131.7251 & 0.4606 & 261.1729 \\
\hline
\end{tabular}
\caption{Optimal Investment Ratio and Max Profit for Short-Period and Long-Period Scenarios}
\label{tab:scenario_result}
\end{table}

In the short term, Immediate decarbonization leads to high profits because it allows the firm to achieve early efficiency gains and cost savings from the rapid adoption of low-carbon technologies. By quickly reducing energy and resource inefficiencies, the firm can capitalize on immediate cost reductions, enhanced operational performance, and potentially favorable regulatory incentives for early movers in decarbonization. These benefits outweigh the initial financial burden of the transition, resulting in the highest short-term profitability. 

However, in the long term, the economic dynamics inevitably shift. The substantial upfront costs associated with Immediate decarbonization, including significant investments in low-carbon technologies and potential disruptions to existing operations, begin to erode long-term profitability. These high initial costs reduce the firm's capacity to sustain investments and adapt to evolving market and regulatory conditions over time. In contrast, the Slow decarbonization scenario allows the firm to distribute its transition costs more evenly over a longer period. This gradual approach minimizes financial strain, enhances the firm’s ability to adapt incrementally, and provides a greater degree of flexibility to optimize investments as low-carbon technologies become matured and more cost-effective. As a result, the Slow strategy enables the firm to achieve a higher long-term profitability. The Quick scenario represents a middle ground between Immediate and Slow strategies. In the short term, its profitability is lower than Immediate decarbonization due to moderate initial costs and slower realization of efficiency gains. However, in the long term, Quick decarbonization surpasses the Immediate scenario in profitability, as it balances higher levels of initial investments with a sufficient degree of flexibility to adapt and sustain low-carbon operations.

Overall, the results highlight the importance of strategic planning in decarbonization efforts. Firms must weigh the trade-offs between short-term efficiency gains and long-term financial stability when determining their optimal transitional investment ratios. While Immediate decarbonization may deliver immediate benefits, a gradual approach such as the Slow strategy offers a more sustainable path to maximizing long-term profits and resilience. This underscores a critical role of proactive, well-paced investment planning in achieving both financial and sustainability goals.

\subsection{Optimal Transition Investment Ratio with Government Intervention}

From the conclusions drawn from the previous subsections, it is evident that companies may exhibit a deficiency in transformational investments in certain scenarios. In such instances, it becomes especially relevant for the government to undertake steps of market adjustments at both macro and micro levels, employing tools such as subsidies or tax penalties. 

Subsidies can serve as a useful incentive for companies by lowering the financial burden associated with making a transition to low-carbon technologies. These subsidies can take various forms, such as direct financial assistance, tax credits, or grants for research and development in green technologies. By reducing the initial costs, subsidies make low-carbon investments more attractive to companies, thereby accelerating the adoption of sustainable practices. For example, a company might receive a tax credit for every unit of renewable energy produced, which would directly offset its production costs and enhance profitability.

On the other hand, tax penalties can act as a deterrent against environmentally harmful practices. By imposing higher taxes on carbon emissions or non-compliant activities, governments can create the risk of a financial disincentive for companies to continue with high-carbon operations. This approach compels businesses to internalize the environmental costs of their actions, thereby encouraging them to seek out more sustainable alternatives. For instance, a carbon tax that increases with the level of emissions will drive companies to innovate and invest in cleaner technologies to minimize their tax liabilities.

Governments have a vested interest in incentivizing companies to undertake low-carbon transitions for several important reasons. Firstly, fostering a transition to low-carbon practices aligns with broader environmental and climate goals, mitigating the adverse effects of climate change and reducing overall carbon emissions in the country. Secondly, by encouraging businesses to adopt sustainable practices, governments contribute to the development of a more resilient and environmentally conscious economy. Additionally, promoting low-carbon transitions enhances a country’s global competitiveness by positioning it as a leader in clean and sustainable technologies. This, in turn, attracts environmentally conscious investments, boosts innovation, and creates green jobs, fostering economic growth. Furthermore, by mitigating environmental risks and promoting sustainability, governments aim at improving public health and reduce the strain on healthcare systems in the country. Incentivizing low-carbon transitions therefore reflects a strategic approach to building a more sustainable and prosperous future, where economic development is harmonized with environmental stewardship.


In the extended model to be presented below, we categorize government regulatory tools into two groups: one from a macro-economic perspective and the other from a micro-economic perspective. The government acts as an `invisible hand'. We assume that from a macro-economic perspective, the government subsidizes carbon prices that are too high in the market and taxes carbon prices that are too low in the market. At a micro-economic level, if a company surpasses a specified threshold for carbon emissions, the government levies taxes. Conversely, if a company's total emissions fall below the threshold, the government offers rewards and subsidies to companies, fostering increased enthusiasm for low-carbon transformation initiatives. To introduce the extension of the the previous model, additional variables are introduced in Table \ref{table3}.

\begin{table}[h]
    \centering
    \begin{tabular}{|c|c|}
    \hline
         Notation& Description\\
         \hline
         $s_1$&  Subsidy Rate for High Carbon Price\\
         $s_2$& Subsidy Rate for Low Carbon Emission\\
         $q_1$& Tax Rate for Low Carbon Price\\
         $q_2$& Tax Rate for High Carbon Emission\\
         $pr_1=Pr(\text { Carbon Price} (B) \geqslant cp)$& Probability of Subsidy for High Carbon Price\\
         $pr_2=Pr(\text { Carbon Emission} \leqslant ce)$& Probability of Subsidy for Low Carbon Emission\\
         \hline
    \end{tabular}
    \caption{Variable Description for the Extended Framework}
    \label{table3}
\end{table}

Based on these variables, the expected total profit of the company after government subsidy or tax can be expressed as:

\begin{equation}
\begin{aligned}
\pi^{\prime}=\pi & +s_1 \cdot (\text { Total Unit Amount }) \cdot p r_1-q_1 \cdot (\text { Total Unit Amount }) \cdot\left(1-p r_1\right) \\
& +s_2 \cdot(\text { Carbon Emission }) \cdot p r_2-q_2 \cdot(\text {Carbon Emission }) \cdot\left(1-p r_2\right) \\
= \pi & +s_1 \cdot \left\{\left(\frac{2}{1+e^{-\alpha k}}\right) \cdot[(1-\alpha) \cdot A]^\beta\right\} \cdot p r_1-q_1 \cdot \left\{\left(\frac{2}{1+e^{-\alpha k}}\right) \cdot[(1-\alpha) \cdot A]^\beta\right\} \cdot\left(1-p r_1\right) \\
& +s_2 \cdot\left(2-\frac{2}{1+e^{-\alpha k}}\right) \cdot\left\{\left(\frac{2}{1+e^{-\alpha k}}\right) \cdot[(1-\alpha) \cdot A]^\beta\right\} \cdot p r_2 \\
& -q_2 \cdot\left(2-\frac{2}{1+e^{-\alpha k}}\right) \cdot\left\{\left(\frac{2}{1+e^{-\alpha k}}\right) \cdot[(1-\alpha) \cdot A]^\beta\right\} \cdot\left(1-p r_2\right) \\
\end{aligned} 
\end{equation}
where, the total profit \(\pi\), excluding government intervention, is calculated as previously shown in Equation (\ref{equ:pi}).


Next, we expand our numerical analysis to explore the impact of government intervention on the optimal transition investment ratio $(\alpha)$ and profitability of the firm, through a revised objective function that incorporates both subsidies and taxes linked to carbon pricing and emissions. The new objective function, $\pi^{\prime}$, integrates additional terms representing financial incentives and penalties for the firm based on carbon-related metrics. 

Our expanded model considers the following variables: subsidy rates ( $s_1$ and $s_2$ ) for high carbon prices and low carbon emissions, respectively; tax rates ( $q_1$ and $q_2$ ) for low carbon prices and high carbon emissions, respectively; and the probabilities ( $p r_1$ and $p r_2$ ) of these rates being applied based on carbon price and emission thresholds. These factors are critical for determining the cost-benefit landscape of environmental compliance and sustainable investment, as they directly influence the incentives and penalties faced by firms.

The baseline variables from our previous studies are maintained in this study for consistency, but now we add more variables to the model. The baseline values for these new variables are shown in Table \ref{table4}. Then we systematically vary each of the new variables individually to observe their effects on the optimal value of $\alpha$ and overall profitability for the firm. Graphical representations will be used to illustrate the relationship between these variables  and the resulting economic outcomes, providing clear insights into the dynamics at play for the firm under different policy scenarios. 

\begin{table}[H]
    \centering
    \begin{tabular}{|c|c|c|}
    \hline
         Notation & Description & Baseline Value\\
         \hline
         $s_1$&  Subsidy Rate for High Carbon Price& 0.8\\
         $s_2$& Subsidy Rate for Low Carbon Emission & 0.8 \\
         $q_1$& Tax Rate for Low Carbon Price & 0.6\\
         $q_2$& Tax Rate for High Carbon Emission &0.6\\
         $pr_1$& Probability of Subsidy for High Carbon Price & 0.5 \\
         $pr_2$& Probability of Subsidy for Low Carbon Emission & 0.5\\
         \hline
    \end{tabular}
    \caption{New Baseline Values for Analysis}
    \label{table4}
\end{table}

\begin{figure}[H]
\includegraphics[width=12cm]{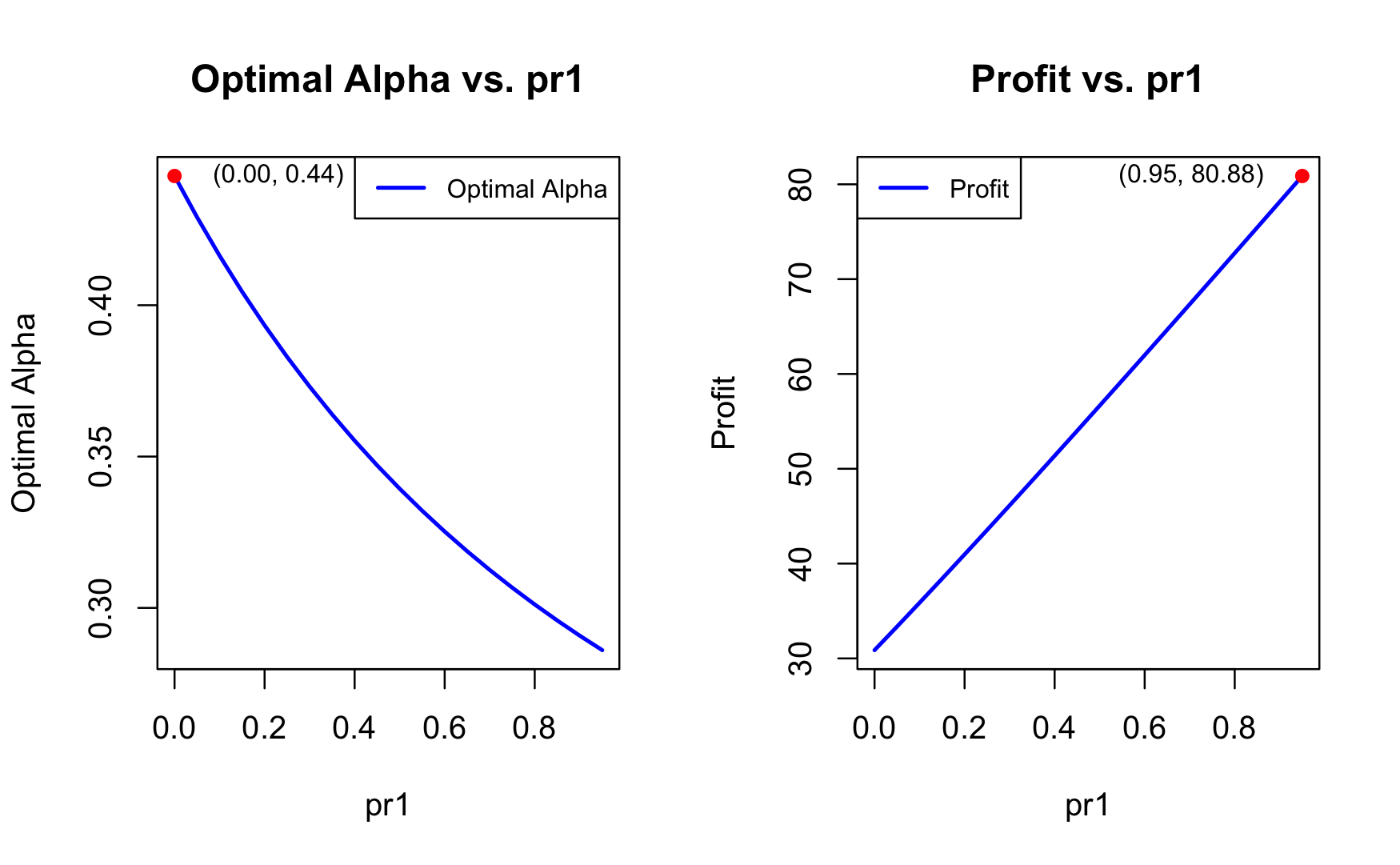}
\centering
\caption{Optimal Level of Alpha and Profit Graph w.r.t. $pr_1$}
\label{figpr1}
\end{figure}

\begin{figure}[H]
\includegraphics[width=12cm]{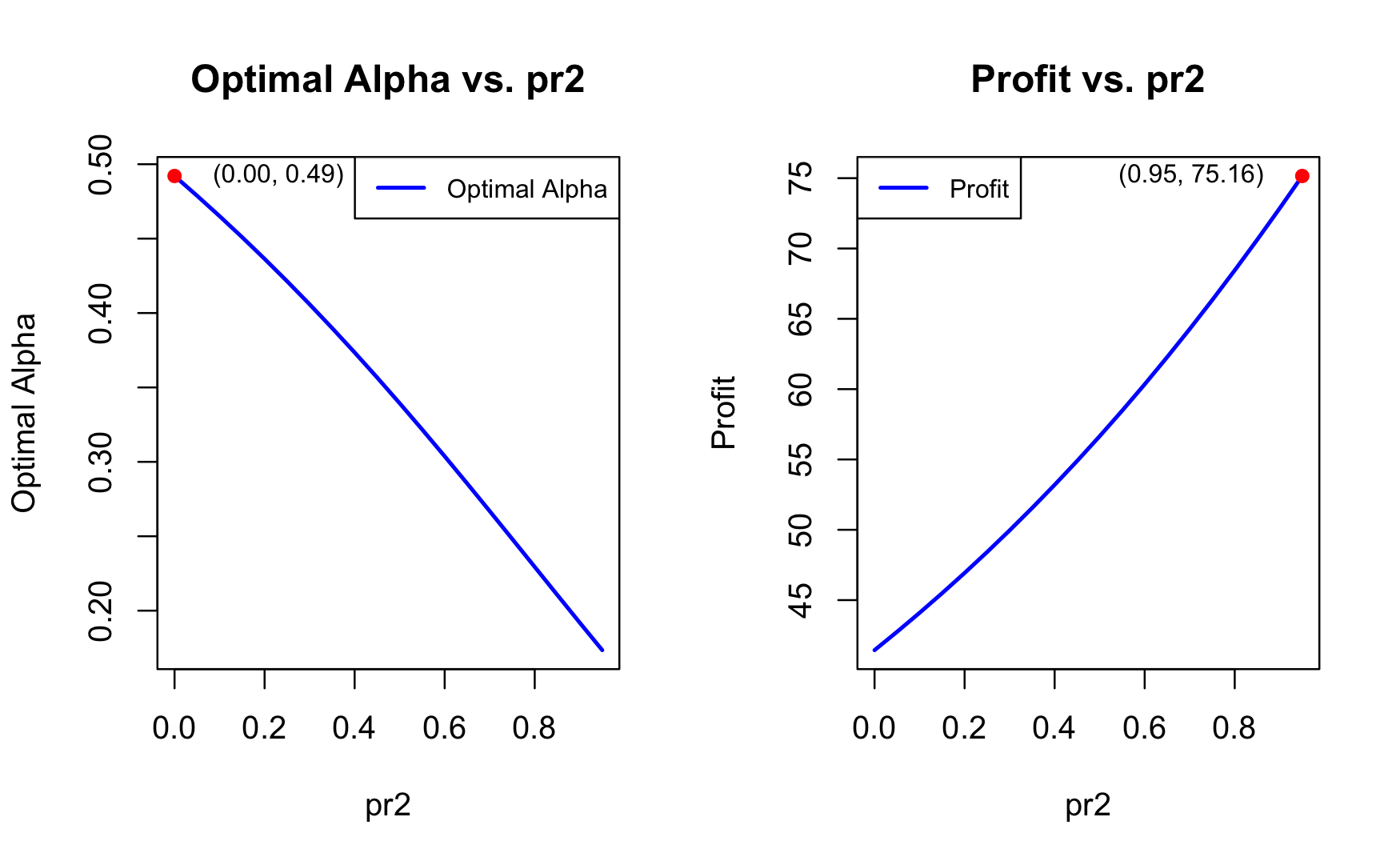}
\centering
\caption{Optimal Level of Alpha and Profit Graph w.r.t. $pr_2$}
\label{figpr2}
\end{figure}

\begin{figure}[H]
\includegraphics[width=12cm]{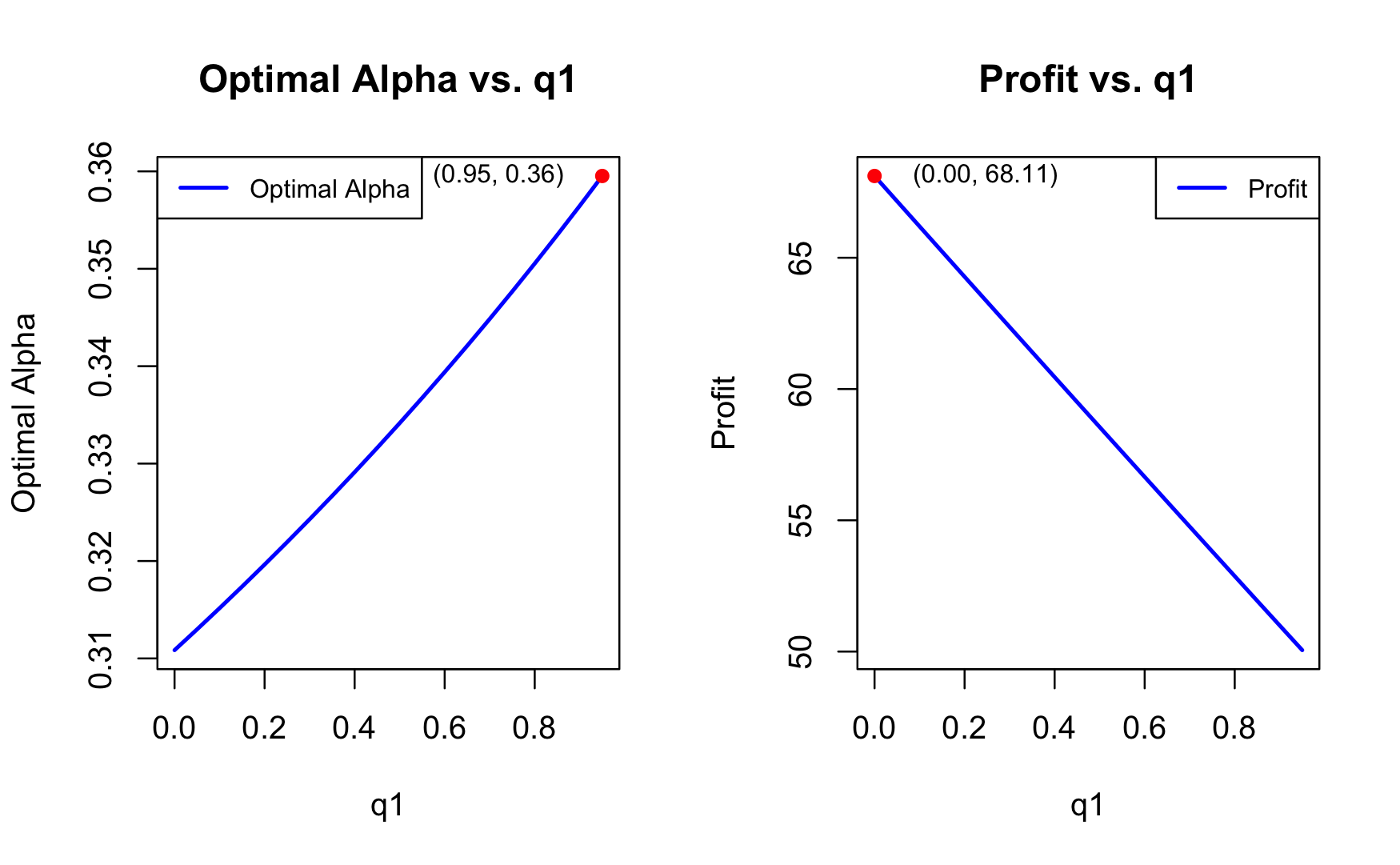}
\centering
\caption{Optimal Level of Alpha and Profit Graph w.r.t. $q_1$}
\label{figq1}
\end{figure}

\begin{figure}[H]
\includegraphics[width=12cm]{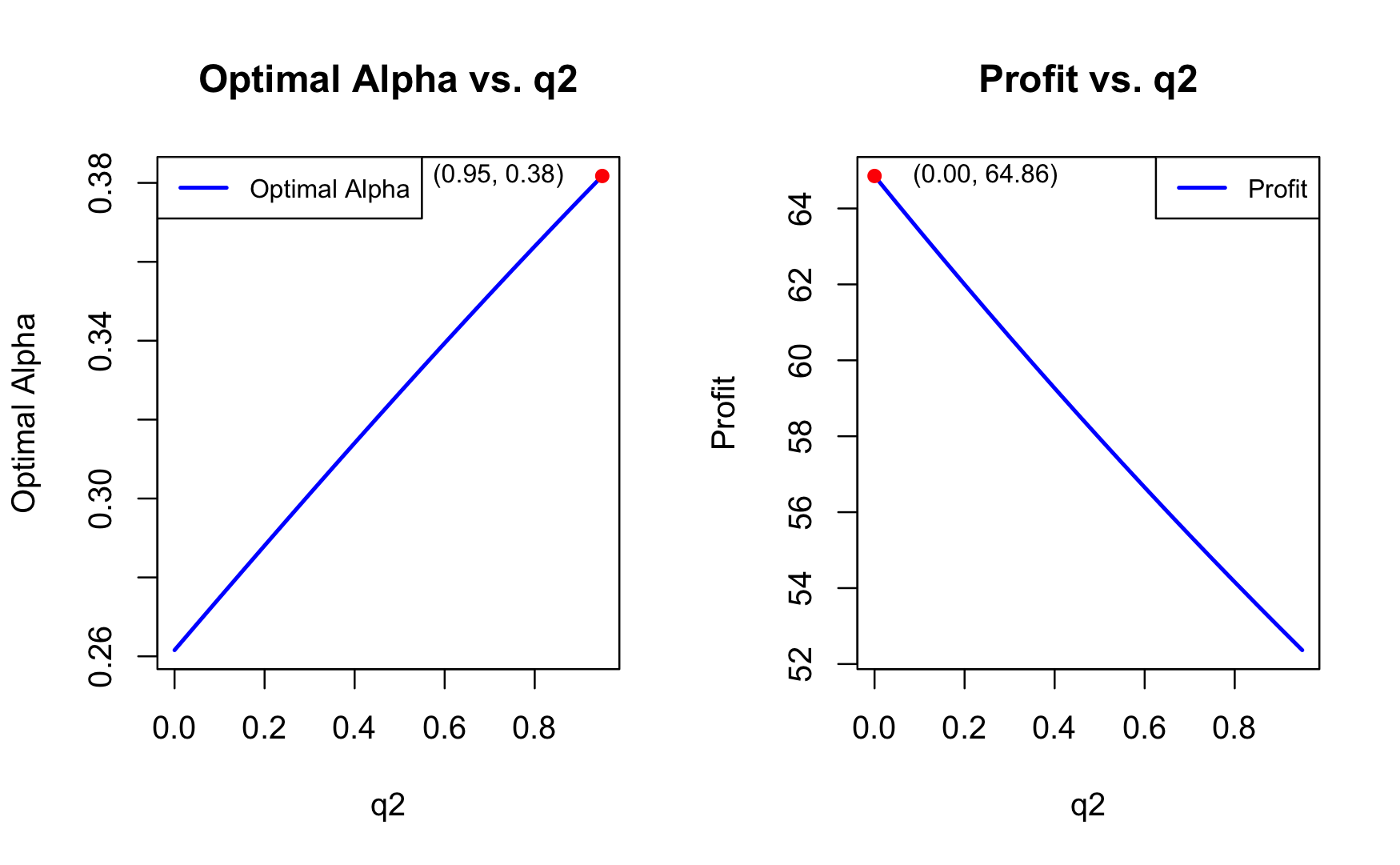}
\centering
\caption{Optimal Level of Alpha and Profit Graph w.r.t. $q_2$}
\label{figq2}
\end{figure}

\begin{figure}[H]
\includegraphics[width=12cm]{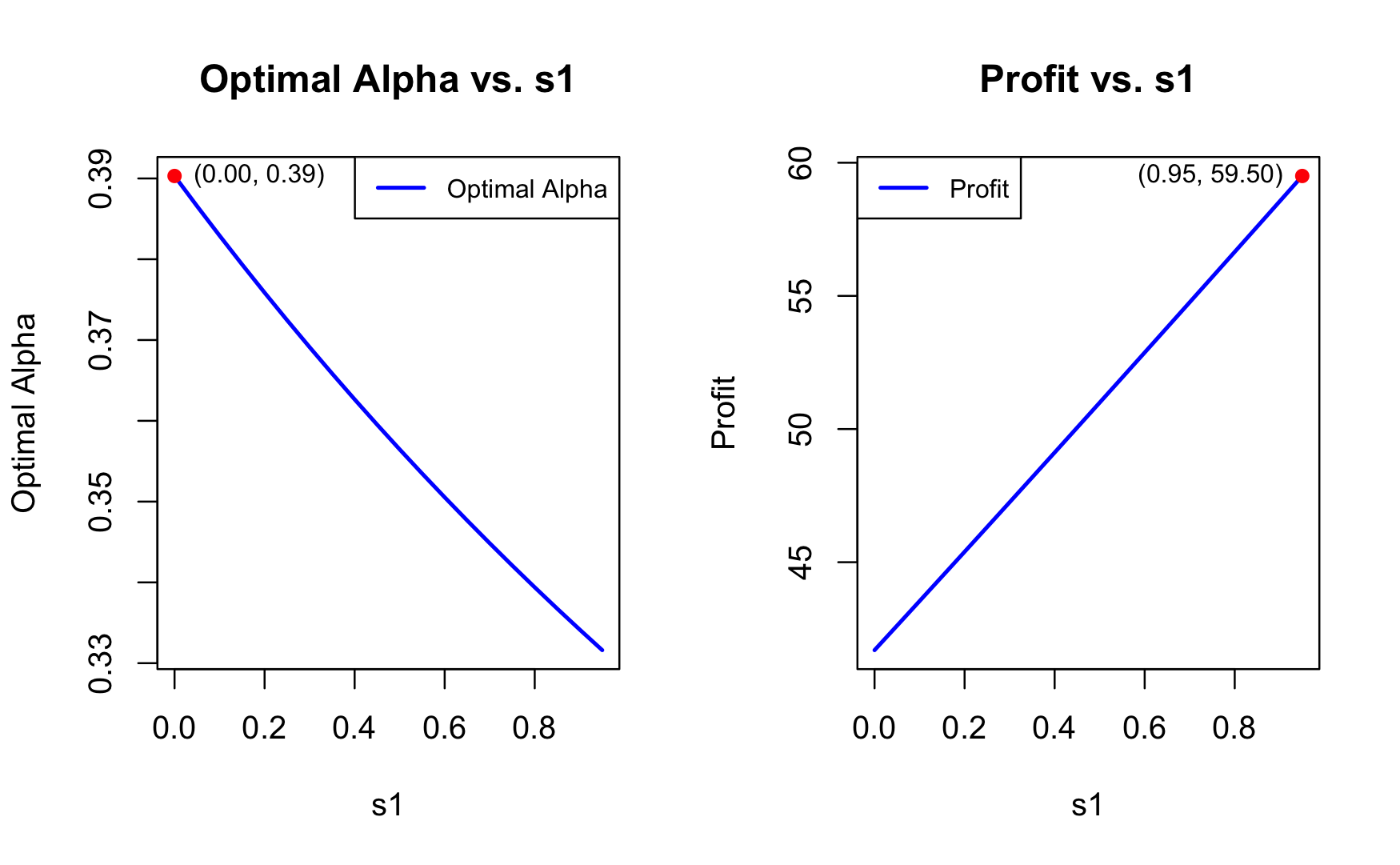}
\centering
\caption{Optimal Level of Alpha and Profit Graph w.r.t. $s_1$}
\label{figs1}
\end{figure}

\begin{figure}[H]
\includegraphics[width=12cm]{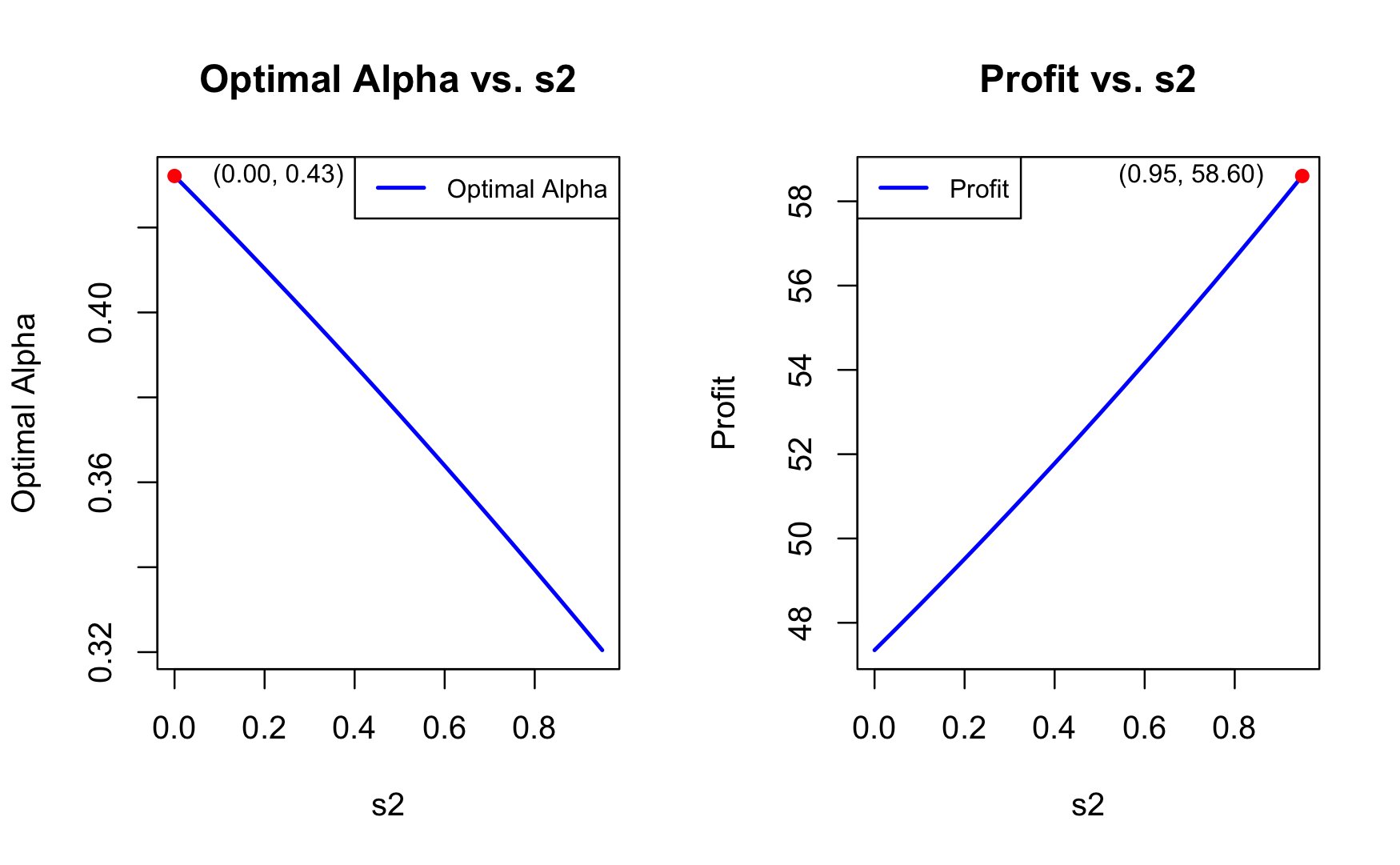}
\centering
\caption{Optimal Level of Alpha and Profit Graph w.r.t. $s_2$}
\label{figs2}
\end{figure}

From Figure \ref{figpr1} to Figure \ref{figs2}, the left plot illustrates how changes in the key variable affect the optimal transition investment ratio with the goal of maximizing profit under each scenario. The right plot shows the corresponding maximum profit achieved by the firm under the optimal transition investment as the key variable changes. We will explain each figure in detail below.

\noindent \textbf{Impact of Probability of High Carbon Price Subsidy (\( p r_1 \)) on Optimal Alpha and Profit (Figure \ref{figpr1})}.
As the probability of receiving a subsidy at higher carbon prices (\( p r_1 \)) increases, the optimal investment ratio (\(\alpha\)) for the firm decreases. This suggests that businesses, when expecting financial support through subsidies, feel less pressure to invest heavily in low-carbon technologies. The security provided by expected subsidies reduces the need for extensive investment by the firm in hedging against future regulatory changes or market conditions that favor lower emissions. This also highlights a strategic point for businesses: relying too heavily on subsidies might provide a short-term relief but could risk an insufficient preparation for a future with stricter environmental regulations. 

Profit increases alongside higher subsidy probabilities, reinforcing the positive impact of government incentives on financial performance of the firm. For investors, this underscores the appeal of firms with strong expectations of a subsidy support, as subsidies directly enhance profitability of the firm. However, it is essential to remain cautious -- companies that over-rely on subsidies may be less motivated to invest in long-term carbon reduction strategies, posing potential risks to the firm as market dynamics or regulations shift.

\textbf{Impact of Probability of Low Carbon Emission Subsidy (\( p r_2 \)) on Optimal Alpha and Profit (Figure \ref{figpr2})}

A rising \( pr_2 \) reflects a higher tolerance from the government for carbon emissions, effectively lowering the threshold required for firms to qualify for subsidies. This shift reduces the incentive for companies to invest in low-carbon transitions, as they can now receive financial support with less stringent emissions standards. Consequently, the optimal value of \(\alpha\), which represents the proportion of resources allocated to low-carbon investments, declines. This trend shows that firms are less motivated to prioritize sustainable transitions when subsidies become more accessible without rigorous emission targets.

On the profitability side, a higher \( pr_2 \) initially appears to be advantageous. With a reduced amount of pressure to invest in transition efforts, firms can allocate fewer resources to low-carbon initiatives, which boosts short-term profits. However, this profit increase may be deceptive, as it comes at the expense of long-term resilience. Firms that minimize transition investments now could face significant financial risks if stricter carbon policies are introduced in the future. A sudden regulatory shift would leave them underprepared to adapt, potentially leading to substantial losses. This underscores a trade-off between short-term profit gains and the uncertain long-term stability that comes with underinvesting in sustainable practices.

\noindent \textbf{Impact of Tax Rates for Low and High Carbon Emissions (\( q_1 \) and \( q_2 \)) on Optimal Alpha and Profit (Figure \ref{figq1} \& Figure \ref{figq2})}.
Higher tax rates on low carbon prices (\( q_1 \)) or high carbon emissions (\( q_2 \)) push the optimal value of \(\alpha\) higher, indicating that businesses are incentivized to adopt cleaner technologies more aggressively when facing heavier tax burdens. This reflects the practical use of taxes as a tool for driving environmental sustainability—businesses react to them by increasing investments in green technologies to mitigate the financial impact of taxes. However, businesses must be strategic at the same time: while aggressive adoption of low-carbon technologies can reduce future tax liabilities for the firm, over-investing without aligning with long-term sustainability goals could lead to a suboptimal resource allocation.

On the flip side, profits decrease as taxes rise, which makes high-emission activities less financially attractive for the firm. This highlights the financial risks for companies slow to adopt cleaner practices, reinforcing the need for a more proactive sustainability planning. Investors should be mindful of firms exposed to significant carbon taxes, as their profitability may suffer without adequate mitigation strategies.

\noindent \textbf{Impact of Subsidy Rates for High and Low Carbon Scenarios (\( s_1 \) and \( s_2 \)) on Optimal Alpha and Profit (Figure \ref{figs1} \& Figure \ref{figs2})}.
An increase in subsidy rates for high carbon price (\( s_1 \)) or low carbon emission (\( s_2 \)) scenarios results in a lower optimal value of \(\alpha\), showing that businesses feel less urgency to invest heavily in green technologies when financial incentives make it easier to comply with emission standards. While this may seem beneficial for companies in the short term, there is a risk that such subsidies could slow down the transition to greener technologies if businesses rely too heavily on them without making the necessary levels of investment in long-term sustainability.

Profit increases with higher subsidies, emphasizing the role of financial support in boosting corporate earnings. From an investment perspective, firms with access to a substantial amount of subsidies for meeting environmental standards are likely to be more profitable in the short term, making them more appealing to investors. However, investors should also assess whether these companies are using the subsidy windfall to fund long-term sustainable transformations, as over-reliance on temporary fiscal incentives could pose challenges for the firm in the face of evolving market or regulatory conditions.

\noindent \textbf{Policy Recommendations for Effective Low-Carbon Transitions}.
Encouraging effective low-carbon transitions requires a sensitive balance between assertive tax policies and a more targeted, limited subsidy approach. The above numerical analysis suggests several important and practical policy recommendations. While subsidies boost short-term profitability, they do not appear to incentivize long-term investment in low-carbon technologies adequately. This is evident from the decreasing trend in the optimal investment ratio (\(\alpha\)) as subsidy probabilities increase. On the other hand, higher emissions taxes consistently drive higher values of \(\alpha\), indicating that tax policies play a more effective role in promoting robust investments in sustainable technologies.

\noindent \textbf{Expanding on Tax Strategies.} A more structured and progressive carbon tax regime could enhance low-carbon transitions more effectively. By gradually increasing the cost of higher emissions, such a policy would not only penalize environmentally harmful practices but also provide businesses with the level of predictability they need in order to plan long-term investments in sustainability. This tax strategy could be further refined by introducing tax incentives for specific technologies that deliver substantial carbon reductions. For businesses, the clear financial penalties for high emissions and rewards for investing in sustainable solutions would create a strong motivation to prioritize clean technology adoption, ensuring a smoother and more predictable transition.

\noindent \textbf{Reassessing Subsidy Schemes.} A more passive and targeted approach to subsidies is recommended. Direct financial incentives should be limited to sectors of the economy where low-carbon alternatives are not yet economically viable or where early-stage technologies need support to scale. This approach minimizes the risk of companies relying on subsidies for profitability without making a genuine progress toward sustainability. It also encourages firms to seek out real, long-term solutions rather than temporary financial relief. Companies should anticipate stricter regulatory standards in the future and take a proactive approach to sustainability. Rather than depending on subsidies alone, businesses are advised to prioritize investments in scalable, low-carbon technologies that align with long-term industry trends. Building internal capabilities for sustainable practices and integrating environmental risk assessments into financial planning can help ensure resilience against future regulatory shifts. Additionally, companies should explore partnerships with other industry players or research institutions to advance low-carbon innovations, as these collaborations can reduce costs and accelerate progress toward sustainability goals.

In summary, integrating fiscal policies, such as carbon taxes and subsidies, with broader regulatory measures can amplify their effectiveness with respect to promoting long-term sustainability and reducing emissions. For instance, mandatory carbon emission disclosures and stricter compliance requirements would not only improve transparency but also make tax and subsidy regimes more impactful. Businesses would be incentivized to adopt greener practices both to avoid penalties and to benefit from tax reliefs. This would create a feedback loop where fiscal and regulatory measures reinforce each other, driving a greater amount of corporate investments in sustainability. 

Additionally, corporate leaders should adopt a forward-thinking approach, moving beyond short-term financial benefits emanating from current subsidies and tax reliefs. Strategic investments in low-carbon technologies today will prepare businesses for future regulatory shifts, reducing the risk of compliance costs down the line. Companies should conduct detailed cost-benefit analyses that balance immediate economic advantages with future regulatory expectations. Proactively enhancing sustainability measures will not only align businesses with global environmental trends but also provide a competitive edge in an increasingly eco-conscious market. This approach places companies in a better position against stricter regulations in the future and boosts their appeal to environmentally aware consumers. 

Finally, investors should closely monitor governmental shifts in environmental policies, as these will impact the profitability and strategic positioning of companies in their portfolios. There is a clear advantage in diversifying investments to include companies with strong environmental strategies or those investing heavily in green technologies. Firms that are better prepared for stringent environmental regulations are more likely to thrive, benefiting from enhanced market reputation and consumer loyalty. Additionally, investors have the potential to influence corporate behavior by advocating for stronger sustainability practices within the companies they support. Active stewardship not only aligns investments with broader environmental goals but also helps mitigate the risks associated with regulatory changes, securing long-term profitability while promoting global sustainability objectives.

\section{Conclusions}

This paper explores the dynamics of firm's strategic investment decisions that support the transition to a low-carbon economy, focusing on the interplay between government policies and corporate strategies. It reveals how corporations adjust their investment strategies based on anticipated returns and environmental impacts, with a cautious approach prevailing in the absence of government incentives. The analysis highlights that, while long-term sustainability goals are beneficial, they may impose significant short-term financial burdens on companies. When government intervention, through subsidies and taxes, is introduced, corporate behavior shifts dramatically. The findings in this paper show that assertive tax policies on carbon emissions effectively drive firm's investment in sustainable technologies, aligning corporate actions with environmental targets. However, while subsidies enhance short-term profitability of firms, they can lead to a greater reliance on government aid and do not necessarily promote sustained innovation in green technologies.

The study suggests a balanced policy approach—where taxes can be used assertively to stimulate investment, and subsidies can be applied selectively to avoid dependency. This would encourage corporations to take proactive steps toward sustainability, aligning with long-term environmental goals rather than relying solely on government incentives. For stakeholders, particularly investors, the research underscores the importance of identifying companies that are actively adapting to evolving fiscal and regulatory landscapes. These companies are better positioned to meet future regulatory demands and capitalize on the growing market emphasis on sustainability, offering better long-term returns.

Extending the one-period model to a multi-period setting under different decarbonization scenarios further reveals how the optimal investment ratio (\(\alpha\)) evolves over time. While aggressive early decarbonization yields the highest short-term profits for the firm, slower approaches offer better long-term profitability due to the lower upfront costs. The analysis also shows that sustainable investment in low-carbon technologies is beneficial in the long run, particularly in scenarios involving gradual transitions. In conclusion, this study provides practical insights for policymakers and business leaders on how to promote investments in low-carbon technologies. By implementing a balanced mix of taxes and targeted subsidies, governments can effectively encourage corporate sustainability efforts.

\bibliographystyle{elsarticle-harv}
\bibliography{main}

\end{document}